\documentclass[aps,longbibliography,pre,reprint,groupaddress,bibnotes]{revtex4-1}

\usepackage{dcolumn}
\usepackage{bm, amssymb, color}
\usepackage[colorlinks=true,linkcolor=blue,citecolor=blue,urlcolor=blue,bookmarks,breaklinks=true]{hyperref}
\usepackage{natbib}
\usepackage{tabulary}
\usepackage{multirow}
\usepackage{amsmath,soul}
\usepackage[export]{adjustbox}
\usepackage{float}
\usepackage[caption=false]{subfig}
\usepackage{array}
\usepackage{cases}
\usepackage{makecell}
\usepackage{physics}
\usepackage{yhmath}
\usepackage{enumitem}

\usepackage[colorinlistoftodos,shadow,textwidth=18mm]{todonotes}

\usepackage{soul,xcolor}
\setstcolor{red}

\usepackage[colorinlistoftodos,shadow,textwidth=18mm]{todonotes}

\begin{document}

\title{Local Number Fluctuations in Hyperuniform and Nonhyperuniform Systems: Higher-Order Moments and Distribution Functions }

\author{Salvatore Torquato}
\email[]{Email: torquato@princeton.edu}
\affiliation{Department of Physics, Princeton University, Princeton, NJ 08544, USA}
\affiliation{Department of Chemistry, Princeton Institute for the Science and Technology of Materials, and Program in Applied and Computational Mathematics, Princeton University, Princeton, New Jersey 08544, USA}

\author{Jaeuk Kim}
\email[]{Email: jaeukk@princeton.edu}
\affiliation{Department of Physics, Princeton University, Princeton, NJ 08544, USA}

\author{Michael A. Klatt}
\email[]{Email: mklatt@princeton.edu}
\altaffiliation[Present address: ]{Friedrich-Alexander-Universit\"at Erlangen-N\"urnberg (FAU), Institut f\"ur Theoretische Physik, Staudtstr. 7, 91058 Erlangen, Germany,
and Department of Experimental Physics, Saarland University, Campus E2 9, 66123 Saarbrücken, Germany}
\affiliation{Department of Physics, Princeton University, Princeton, NJ 08544, USA}

\date{\today}

\begin{abstract}

The local number variance $\sigma^2(R)$ associated with a spherical sampling window of radius $R$ enables a classification of many-particle systems in $d$-dimensional Euclidean space $\mathbb{R}^d$ according to the degree to which large-scale density fluctuations are suppressed, resulting in a demarcation between  hyperuniform and nonhyperuniform phyla.
To more completely characterize density fluctuations, we carry out an extensive study of  higher-order moments or cumulants, including the {\it skewness} $\gamma_1(R)$, {\it excess  kurtosis} $\gamma_2(R)$ and the corresponding probability distribution function $P[N(R)]$ of a large family of models across the first three space dimensions, including both hyperuniform and nonhyperuniform systems with varying degrees of short- and long-range order.
To carry out this comprehensive program, we derive new theoretical results that apply to general point processes and conduct high-precision numerical studies.
Specifically, we derive explicit closed-form integral expressions for $\gamma_1(R)$ and $\gamma_2(R)$ that encode structural information up to three-body and four-body correlation functions, respectively.
We also derive rigorous bounds on $\gamma_1(R)$, $\gamma_2(R)$ and $P[N(R)]$ for general point processes and corresponding exact results for general packings of identical spheres.
High-quality simulation data for $\gamma_1(R)$, $\gamma_2(R)$ and $P[N(R)]$ are generated for each model.
We also ascertain the proximity of  $P[N(R)]$ to the normal distribution 
via a novel  Gaussian ``distance" metric $l_2(R)$. Among all models, the convergence to a central limit theorem (CLT) is generally fastest for the disordered hyperuniform 
processes such that $  \gamma_1(R)\sim l_2(R) \sim R^{-(d+1)/2}$ and $\gamma_2(R)\sim R^{-(d+1)}$ for large $R$. 
The convergence to a CLT is slower for standard nonhyperuniform models and slowest for the ``antihyperuniform" model studied here. We 
prove that  one-dimensional hyperuniform systems of class I or any $d$-dimensional lattice cannot obey a CLT.
Remarkably, we discovered that the gamma distribution provides a good approximation to   $P[N(R)]$ for all  models that obey a CLT
across all dimensions for intermediate to large values of $R$, enabling us  to estimate  
the large-$R$ scalings of $\gamma_1(R)$,  $\gamma_2(R)$ and $l_2(R)$. 
For any $d$-dimensional model  that ``decorrelates" or ``correlates" with $d$, we  elucidate why  $P[N(R)]$ increasingly moves toward or away from Gaussian-like behavior, respectively.
Our work elucidates the fundamental importance of higher-order structural information to  fully characterize density fluctuations in many-body systems across length scales and dimensions, and thus has broad implications for condensed matter physics, engineering, mathematics and biology.
\end{abstract}

\maketitle

\section{Introduction}

The quantification of density fluctuations in many-particle systems in $d$-dimensional Euclidean space $\mathbb{R}^d$ 
is of great fundamental and practical importance in many fields across the
physical, mathematical and biological sciences \cite{Sc66,Ve75,Zi77,Ca78,Ha86,Jo91,Pe93,Ble93,Tr98b,To00a,Ga02,Wa02,To03a,La03,Kl16,To18a}.
It is well known that {\it long-wavelength} density fluctuations of disordered as well as ordered 
systems contain crucial information about the structure as well as equilibrium and nonequilibrium physical properties of the systems \cite{Sc66,Ha86,To03a,To18a}.
Clearly, density fluctuations that occur on some arbitrary
local length scale \cite{Ve75,Zi77,Ble93,Tr98b, Ro99b, Ga02,Wa02,To03a,Kl16,To18a,Zh20c} provide considerably more information about the system than 
those in the long-wavelength limit.

\begin{figure}[H]
\centering{\includegraphics[width = 0.45\textwidth]{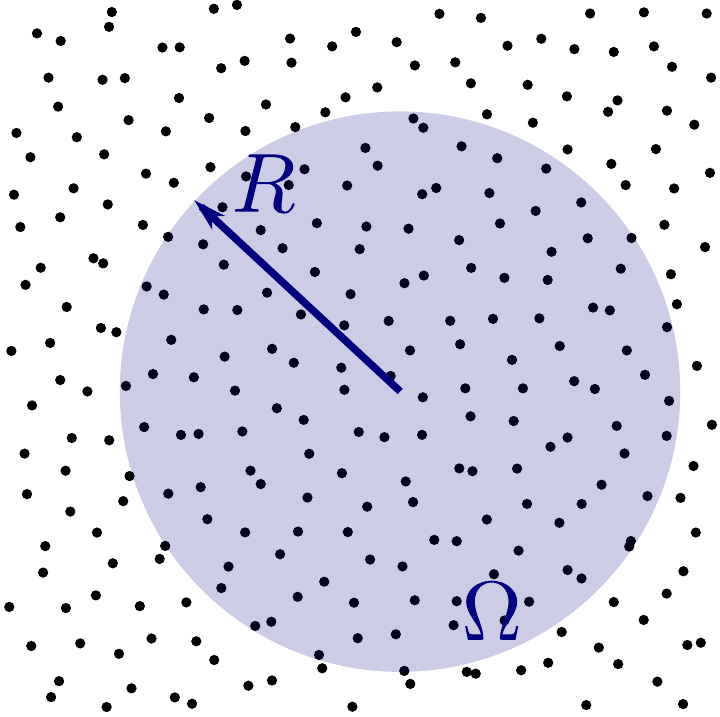}}
\caption{Schematic showing  a spherical window $\Omega$ of radius $R$ used to sample fluctuations in the number of points in
a two-dimensional point process.}
\label{schematic}
\end{figure}

Consider a statistically homogeneous (translationally invariant) point process in $d$-dimensional Euclidean space $\mathbb{R}^d$
at number density $\rho$ and sampling for the number of points $N(R)$ within a $d$-dimensional spherical window of radius $R$ (see Fig. \ref{schematic})
and volume
\begin{equation}
v_1(R)= \frac{\pi^{d/2} R^d}{\Gamma(1+d/2)}.
\label{v1}
\end{equation}
The local number variance $\sigma^2(R)\equiv \langle N^2(R) \rangle  -\langle N(R) \rangle^2 $ is a useful measure of 
number fluctuations, where the first moment $\langle N(R)\rangle =\rho v_1(R)$ is the average number of points within a 
$d$-dimensional spherical (sampling) window of radius $R$ and angular brackets denote
an ensemble average.
 The local number variance is exactly determined by pair statistics, and can be given either in terms of the pair correlation
function $g_2({\bf r})$ in direct space or the structure factor $S({\bf k})$ in reciprocal space \cite{To03a}: 
\begin{eqnarray}
\sigma^2(R) &=& \rho  v_1(R) \left[ 1 +  \rho \int_{\mathbb{R}^d} h({\bf r}) \alpha_2(r;R) \,  d{\bf r} 
\right], \label{local1}\\
&=& \rho  v_1(R)\Big[\frac{1}{(2\pi)^d} \int_{\mathbb{R}^d} S({\bf k}) 
{\tilde \alpha}_2(k;R) d{\bf k}\Big],
\label{local2}
\end{eqnarray}
where  $h({\bf r})\equiv g_2({\bf r})-1$ is the total correlation function, $\alpha_2(r;R)$  is the intersection volume of two spherical windows  of radius $R$, scaled by $v_1(R)$,
whose centers  are separated by the distance $r$, and ${\tilde \alpha}_2(k;R)$ is its Fourier transform.
The large-scale behavior of the number variance  $\sigma^2(R)$
is central to the hyperuniformity concept, which is attracting attention
across many fields \cite{To03a,To18a,Ch18,Zh20b,Sh20,Wi20,Ma20a}. Specifically, a hyperuniform point process is one
in which $\sigma^2(R)$ grows slower than the window volume, i.e., $R^d$, for large
$R$ and hence is characterized by  large-scale density fluctuations that are anomalously suppressed
compared to those of typical disordered systems. The hyperuniformity concept generalizes the traditional notion of long-range order of crystals and quasicrystals to also encompass certain 
exotic disordered states of matter \cite{To03a,To18a}.
Disordered hyperuniform systems are diametrically  opposite to systems at thermal critical points in which the local 
variance diverges faster than $R^d$ in the limit $R \to \infty$.
Any system with such divergent behavior in the local variance has been called {\it anti-hyperuniform}  \cite{To18a} (see also
Sec. \ref{def} for additional details).

While the local number variance contains useful information, one would like 
to more completely characterize the fluctuations by ascertaining  higher-order moments, i.e., $ \langle N^k(R) \rangle $, where
$k\ge 3$, as well as the corresponding discrete probability distribution  $P[N(R)]$
associated with finding exactly $N(R)$
particles within a  $d$-dimensional spherical window of radius $R$.
The $m$th moment of the distribution is given by
\begin{equation}
\langle N^m(R) \rangle =\sum_{N(R)=0} N^m(R) P[N(R)].
\label{mom-P}
\end{equation}
Due to the fact that the random variable $N(R)$ is discrete and cannot take on negative values, 
 the probability distribution $P[N(R)]$, for finite $R$, can never exactly attain the 
normal distribution, which is given by
\begin{equation}
P[N(R)]=\frac{1}{\sqrt{2\pi} \sigma(R)} \exp[\frac{(N(R)-\langle N(R)\rangle)^2}{2\sigma^2(R)}].
\label{G}
\end{equation}
For example, for a statistically homogeneous Poisson point process in $\mathbb{R}^d$ at number density $\rho$,
\begin{equation}
  P[N(R)]= \frac{[\rho v_1(R)]^{N(R)}}{N(R) !} \exp[-\rho v_1(R)],
\label{P}
\end{equation}
which deviates significantly from the normal distribution for sufficiently small $R$.
This is one of the rare cases in which a closed-form analytic formula for $P[N(R)]$ is known across dimensions for nontrivial point processes.
It is only when $R$ tends to infinity that the Poisson distribution
becomes a normal distribution, i.e., it follows a central limit theorem (CLT); see Ref. \cite{La17} and references therein.
The reader is referred to Refs. \cite{Pe02,He06,Sc13,Bl19} for proofs of CLT for other point processes.

In this paper, we investigate the {\it skewness} $\gamma_1(R)$ (related to the first three moments),
{\it excess  kurtosis} $\gamma_2(R)$ (related to the first four moments), and the number distribution function $P[N(R)]$
for general homogeneous point processes in $\mathbb{R}^d$ as well as a wide class of models across dimensions.
Specifically, we  derive explicit closed-form integral expressions for  $\gamma_1(R)$
in terms of the number density $\rho$, the pair correlation function $g_2$ and the three-body
correlation function $g_3$ (defined in Sec. \ref{def}). Similarly, we derive corresponding formulas 
for the  {\it excess  kurtosis} $\gamma_2(R)$, which now depends additionally
on the four-body correlation function $g_4$. These integral relations also involve geometrical
information about the spherical windows via the intersection volumes of up to
three and four spheres in the cases of the skewness and excess kurtosis, respectively. 
Thus, the skewness and excess kurtosis encode up to three-body and four-body
information about spatial correlations and window geometries, respectively.
We  also derive some exact elementary results for the skewness, excess kurtosis and number distribution
that apply to general packings of identical spheres. 

Via high-precision computer simulation studies, we accurately determine the number variance, skewness, excess kurtosis and the number
distribution for up to eight different models of statistically homogeneous point processes across the first
three space dimensions and wide range of window radii $R$. These models include both nonhyperuniform and hyperuniform systems with
varying degrees of short- and long-range order. Among the nonhyperuniform point
processes, we characterize fluctuations of Poisson, random sequential addition (RSA) packings, equilibrium hard spheres,
Poisson cluster and  anti-hyperuniform  point processes. Among the hyperuniform
point processes, we characterize hypercubic lattices, randomly perturbed lattices,
and stealthy disordered hyperuniform systems. We show that our simulation results for  $\gamma_1(R)$, $\gamma_2(R)$ and $P[N(R)]$
for all models are in excellent agreement with the aforementioned rigorous bounds and exact
results for the applicable ranges of $R$. For all disordered hyperuniform models, our explicit general formulas 
of these quantities in terms of $n$-body information enable us to infer
the existence of ``hidden" order that manifests itself for the first time 
at the three-body level or higher.

For each model considered in this paper, we are interested in ascertaining how 
large $R$ must be  such that $P[N(R)]$ is well-approximated by the Gaussian (normal) distribution.
We have found that such  ``distance" metrics proposed previously are not adequate
for assessing the diverse set of models that we consider here across dimensions.
We quantify this proximity to the normal distribution for any model  by introducing a
certain Gaussian ``distance" metric $l_2(R)$, defined in Sec. \ref{results}. This distance
metric enables us to accurately determine when the distribution function $P[N(R)]$
for a particular model is tending to a CLT. Because the distributions for all
models  (except the lattices) across dimensions are {\it unimodal}, the tendency
to a CLT corresponds to the skewness and excess kurtosis simultaneously tending to zero.
We have found that almost all of the considered models across dimensions obey a CLT.
The  convergence to a CLT is slowest for the anti-hyperuniform point process, 
followed by the Poisson cluster process, and  the Poisson process. The nonhyperuniform
RSA and equilibrium packings tend to a CLT at the same rate as a Poisson point process
but with smaller coefficients of proportionality.
Among all models, the convergence to a CLT is generally fastest for the disordered hyperuniform 
processes. The only models considered that do not achieve a CLT are the hypercubic lattices
for any $d$ and 1D hyperuniform systems of class I. The reader is referred to Sec. \ref{results} for details.

We have examined a variety of well-known closed-form probability distributions $P[N(R)]$
to ascertain those which  best approximate the actual distributions for finite $R$ for all
of our models. Interestingly, the gamma 
distribution provides a good approximation to the number distribution 
$P[N(R)]$ for all  models that obey a CLT
across all dimensions for intermediate to large values of $R$ (Sec. \ref{gamma}). It is noteworthy that  the approximation of $P[N(R)]$ by a gamma distribution 
enables us to estimate  the large-$R$ scalings of $\gamma_1(R)$,  $\gamma_2(R)$ and $l_2(R)$ for all models across 
dimensions that obey a CLT. Among all models, the convergence to a CLT is generally fastest for the disordered hyperuniform 
processes such that $\gamma_1(R)\sim l_2(R) \sim R^{-(d+1)/2}\text{~and~}\gamma_2(R)\sim R^{-(d+1)}$ for large $R$. 
For standard nonhyperuniform models, convergence to a CLT is slower such that 
$  \gamma_1(R)\sim  l_2(R) \sim R^{-d/2}\text{~and~}\gamma_2(R)\sim R^{-d}$. Finally, convergence to a CLT is slowest
for the antihyperuniform  model such that $\gamma_1(R)\sim l_2(R) \sim R^{-1/2}\text{~and~}\gamma_2(R)\sim R^{-1}$.
These predictions are corroborated by corresponding simulation results.

An important fundamental question is what is  the effect of increasing the space dimension
on number fluctuations for any particular model?
To answer this question, we recall the so-called {\it decorrelation principle} \cite{To06b},
which roughly states that for any disordered point process,  unconstrained correlations that exist
in low dimensions vanish as $d$ tends to infinity, and all higher-order correlation functions
$g_n$ for $n \ge 3$ may be expressed in terms of the number density $\rho$ and pair correlation function $g_2$.
The decorrelation principle was employed to justify the conjecture that
the densest sphere packings in sufficiently high dimensions are disordered (as opposed
to ordered in low dimensions) \cite{To06b,Sc08}. Importantly, decorrelation in pair statistics has been shown
to manifest itself in low dimensions in the case of disordered sphere packings  \cite{Sk06,To06a,To06d}
as well as other disordered systems with strongly repulsively interacting  particles \cite{To08b,Za08}. 
Since the number distribution function $P[N(R)]$ generally involves certain integrals
over all of the $n$-body correlation functions, the decorrelation principle implies that $P[N(R)]$ increasingly becomes Gaussian-like as the space dimension increases
for any  model that decorrelates with $d$.
Similarly, for models the correlate with $d$, $P[N(R)]$ increasingly deviates from the normal distribution
as $d$ increases. We confirm  such behaviors for all  models that obey a CLT (Sec. \ref{decorrelate}).

In Sec. II, we provide basic definitions and necessary background material.  In Sec. III, we
derive explicit closed-form integral expressions for  $\gamma_1(R)$ and $\gamma_2(R)$
as well as rigorous lower bounds on both of these quantities. We also
obtain some general exact results for the first few cumulants and distribution
functions for sphere packings, whether disordered or not. Section \ref{models} describes
the large variety of nonhyperuniform and hyperuniform models in one, two and three
dimensions that we study in this paper. In Sec. \ref{distance}, we discuss our proposed Gaussian distance metric.
Section \ref{sec:sampling} describes the simulation procedure
that we employ to sample the first four cumulants and number distributions.
In Sec. \ref{results}, we present our results.
 We make concluding remarks in Sec. \ref{conclusions}.

\section{Definitions and Background}
\label{def}

A stochastic point process  in $\mathbb{R}^d$
is defined as a mapping from a probability space
to configurations of points ${\bf r}_1, {\bf r}_2, {\bf r}_3\ldots$
in $d$-dimensional Euclidean space $\mathbb{R}^d$; see Ref. \cite{Chi13} for mathematical details.
Let $X$ denote the set of configurations such that each configuration
$x \in X$ is a subset of $\mathbb{R}^d$ that satisfies two regularity
conditions: (i) there are no multiple points
(${\bf r}_i \neq {\bf r}_j$ if $ i\neq j$) and (ii)
each bounded subset of $\mathbb{R}^d$ must contain
only a finite number of points of $x$ (i.e.,, $x$ is ``locally finite'').
The point process is statistically
is characterized by the {\it generic} $n$-particle probability density  function $\rho_n({\bf r}^n)$, where ${\bf r}^n$ is a shorthand
notation for the position vectors of any $n$ points, i.e., ${\bf r}^n\equiv {\bf r}_1,{\bf r}_2,\ldots,{\bf r}_n$ \cite{Ha86,To02a}.
In words, the quantity $\rho_n({\bf r}^n)d{\bf r}^n$ is proportional to the probability of finding {\it any} $n$ particles 
with configuration $\bf r^n$ in volume element $d{\bf r}^n\equiv d{\bf r}_{1} d{\bf r}_{2} \cdots  d{\bf r}_n$, i.e.,
it is the {\it probability measure}.
For any subvolume $\Omega \in \mathbb{R}^d$, the following
normalization (average) condition involving the fluctuating number of particles
within this subvolume, $N_\Omega$, immediately follows
\begin{eqnarray}
\left\langle \frac{N_\Omega!}{(N_\Omega -n)!}\right\rangle_{\Omega}=  \int_\Omega  \int_\Omega  \cdots  \int_\Omega \rho_n({\bf r}^n) d{\bf r}^{n},
\label{rhon}
\end{eqnarray}
Note that this random setting is quite general; it incorporates cases in which the locations of the points
are deterministically known, such as a lattice.

For statistically homogeneous media, $\rho_{n}({\bf r}^n)$
is translationally invariant and hence depends only on the relative
displacements, say with respect to ${\bf r}_1$:
\begin{equation}
\rho_{n}({\bf r}^n)=\rho_{n}({\bf r}_{12},{\bf r}_{13},\ldots,{\bf r}_{1n}),
\end{equation}
where ${\bf r}_{ij}={\bf r}_j-{\bf r}_i$. In particular, the one-particle
function $\rho_1$  is just equal to the constant {\it number density} of particles $\rho$. 
For statistically homogeneous point patterns, it is convenient to define the so-called
{\it $n$-particle correlation function}
\begin{equation}
g_n({\bf r}^n) = \frac{\rho_n({\bf r}^n)}{\rho^n}.
\label{nbody}
\end{equation}
In systems without long-range order and in which the
particles are mutually far from one another, $\rho_n({\bf r}^n) \rightarrow \rho^n$ and we have
from (\ref{nbody}) that
$g_n({\bf r}^n) \rightarrow 1$.
Thus, the deviation of $g_n$ from unity  provides a measure of the degree of spatial correlation
between the particles, with unity corresponding to no spatial correlation.

The important two-particle quantity $g_2({\bf r}_{12})$
is usually referred to as the {\it pair correlation function}.
The {\it total correlation function} $h({\bf r}_{12})$ is defined as
\begin{equation}
h({\bf r}_{12})=g_2({\bf r}_{12})-1,
\label{total}
\end{equation}
and thus is a function that is zero when there are no spatial
correlations in the system.
Observe that the structure factor $S(\bf k)$ is related to the Fourier
transform of $h(\bf r)$, denoted by ${\tilde h}({\bf k})$, via the expression
\begin{equation}
S({\bf k})=1+\rho {\tilde h}({\bf k}),
\label{factor}
\end{equation}
where 
\begin{equation}
{\tilde h}({\bf k})=\int_{\mathbb{R}^d} h({\bf r}) e^{-i {\bf k} \cdot {\bf r}}\, d{\bf r}.
\end{equation}
A hyperuniform point process is one in which single-scattering events
at infinite wavelength vanishes, i.e.,
\begin{equation}
\lim_{|{\bf k}|\to 0} S({\bf k})=0.
\label{S-hyp}
\end{equation}
This implies that a hyperuniform system obeys the following sum rule:
in direct space
\begin{equation}
\rho \int_{\mathbb{R}^d} h({\bf r}) \, d{\bf r} =-1,
\end{equation}
and hence $h({\bf r})$ must exhibit negative pair correlations, i.e.,  {\it anticorrelations}, for some values of $\bf r$ \cite{To18a}.
By contrast, an {\it anti-hyperuniform} point process is one in which $S({\bf k})$ tends to $+\infty$
in the limit $|\bf k| \to 0$ \cite{To18a}.

A {\it lattice} $\Lambda$ in $\mathbb{R}^d$ is a subgroup
consisting of the integer linear combinations of independent vectors that span $\mathbb{R}^d$ and thus represents a special subset of point processes.
In a lattice $\Lambda$, the space $\mathbb{R}^d$ can be geometrically divided into identical regions $F$ called {\it fundamental cells}, each of which contains
just one point  specified by the {\it lattice vector} \cite{Co93,To10d}.
In the physical sciences, a lattice is equivalent
to a {\it Bravais} lattice. Unless otherwise stated, we will
use the term lattice. 
A {\it periodic} point process is a
more general notion than a lattice because it is
is obtained by placing a fixed configuration of $N$ points (where $N\ge 1$),
called the {\it basis},
within one fundamental cell of a lattice $\Lambda$, which
is then periodically replicated. Thus, the point process is still
periodic under translations by $\Lambda$, but the $N$ points can occur
anywhere in the chosen fundamental cell.
Any lattice or periodic point configuration can be made statistically homogeneous
by uniform translations of the pattern within the fundamental cell.

We call a {\it packing} in $\mathbb{R}^d$ a collection of nonoverlapping particles
\cite{To18b}. The centroids of the particles constitute a special point process
in which no two particles can closer than some minimal distance. In this
paper, we consider packings of identical spheres of diameter $D$. The packing
fraction $\phi=\rho v_1(D/2)$ is the fraction of space covered by the spheres,
where $v_1(R)$ is the volume of a sphere of radius $R$ given by (\ref{v1}).
Any periodic point configuration with a finite basis can be regarded
to be a packing since there is a minimal pair distance.

\section{General Local Moment Formulas and Probability Distribution for Homogeneous Point Processes}
\label{general}

We consider the determination of local moment formulas using the formalism of Torquato and Stillinger \cite{To03a} 
that was used to obtain formulas for the 
local number variance. Here we immediately begin with a 
$d$-dimensional spherical window of radius $R$ in $d$-dimensional Euclidean space $\mathbb{R}^d$ 
with window indicator function 
\begin{equation}
w({\bf r}-{\bf x}_0;{R})=
\begin{cases}
  1, & \quad {|{\bf r} - {\bf x}_0|} \le R,\\
  0, & \quad {|{\bf r} -{\bf x}_0 |} > R
\end{cases}
\label{window}
\end{equation}
where $\bf x $ is some arbitrary position vector in $\mathbb{R}^d$ and ${\bf x}_0$ is the position vector of the window center.
The number of points ${N}(R;{\bf x}_0)$ within the window  at position ${\bf x}_0$ is given by
\begin{eqnarray}
N(R;{\bf x}_0)&=& \sum_{i=1} w({\bf r}_i-{\bf x}_0;{R}),
\label{N}
\end{eqnarray}
which must be a finite number.
We will subsequently use the fact that
\begin{equation}
v_n^{\mbox{\scriptsize int}}({\bf r}^n;R)= \int_{\mathbb{R}^d} \prod_{i=1}^n w({\bf r}_i-{\bf x}_0; R) d{\bf x}_0
\label{v_n}
\end{equation}
is the intersection volume of $n$ spheres of radius $R$ centered at positions ${\bf r}_1,{\bf r}_2,\cdots,{\bf r}_n$.

\subsection{Moments}

The $m$th moment associated with the random variable $N(R)$
is given by the following ensemble average:
\begin{align}
\langle N^m(R) \rangle &= \Bigg\langle \sum_{i_1=1} \sum_{i_2=1} \cdots \sum_{i_m=1} w({\bf r}_{i_1}-{\bf x}_0;{R})  w({\bf r}_{i_2}-{\bf x}_0;{R}) \nonumber \\
&\cdots\times w({\bf r}_{i_m}-{\bf x}_0;{R}) \Bigg\rangle.
\label{N-m}
\end{align}
Here we have implicitly assumed homogeneity, which renders the $m$th moment independent of the position of the window ${\bf x}_0$.
Under the {\it ergodic} assumption, the ensemble average indicated on the left-hand side
of relation (\ref{N-m}) is equivalent to averaging by uniformly window sampling a single realization
over the infinite space.

Following the same procedure used in Ref. \cite{To03a} to obtain an explicit formula for the  second moment,
we obtain from (\ref{N-m}) that the $m$th moment $\langle N^m(R) \rangle$ for a homogeneous process is given by integrals involving the finite set of correlation functions $g_2,g_3,\ldots g_m$ weighted with the set of intersection volumes $v_2^{\mbox{\scriptsize int}}, v_3^{\mbox{\scriptsize int}},\ldots, v_m^{\mbox{\scriptsize int}}$.
For example, expanding the sums in (\ref{N-m}) into one-body, two-body, $\ldots$ and $m$-body terms
in a manner analogous to the one given in Ref. \cite{To03a}, third and fourth moments
are explicitly given by
\begin{eqnarray}
\langle N^3({R})\rangle
&=&\rho v_1(R) +3\rho^2\int_{\mathbb{R}^d}  g_2({\bf r}^3)
   v_2^{\mbox{\scriptsize int}}(r;R) d{\bf r}\nonumber \\
&& +\, \rho^3\int_{\mathbb{R}^d}  \int_{\mathbb{R}^d} g_3({\bf r}^3) 
v_3^{\mbox{\scriptsize int}}({\bf r}^3;R) d{\bf r}_2 d{\bf r}_3,
\label{N3}
\end{eqnarray}
and
\begin{align}
\langle N^4({R}) \rangle &=\rho v_1(R) +7\rho^2\int_{\mathbb{R}^d}  g_2({\bf r})
v_2^{\mbox{\scriptsize int}}(r;R) d{\bf r} \nonumber \\
 + &6\rho^3\int_{\mathbb{R}^d}  \int_{\mathbb{R}^d} g_3({\bf r}^3)
 v_3^{\mbox{\scriptsize int}}({\bf r}^3;R)  d{\bf r}_2 d{\bf r}_3 \nonumber \\
  + &\rho^4\int_{\mathbb{R}^d}  \int_{\mathbb{R}^d}  \int_{\mathbb{R}^d} g_4({\bf r}^4)
v_4^{\mbox{\scriptsize int}}({\bf r}^4;R)  d{\bf r}_2 d{\bf r}_3 d{\bf r}_4 ,
\label{N4} 
\end{align}
where $v_n^{\mbox{\scriptsize int}}$ is given by (\ref{v_n}).

\begin{widetext}
We are generally interested in the $m$th order {\it cumulant} $C_m(R)$, which is
directly related to the  $m$th {\it central moment}  $\langle [N(R) - \langle N(R) \rangle]^m\rangle$.
For example, the first several cumulants are given by
\begin{eqnarray}
C_2(R)&\equiv&  \sigma^2(R)\nonumber \\
&=& \rho v_1(R) +\rho^2\int_{\mathbb{R}^d}  h({\bf r})
v_2^{int}(r;R) d{\bf r} \nonumber \\
&=& \rho v_1(R)[1-\rho v_1(R)] +\rho^2\int_{\mathbb{R}^d}  g_2({\bf r})
v_2^{int}(r;R) d{\bf r},
\label{C2}
\end{eqnarray}
\begin{eqnarray}
\hspace{-0.6in}C_3(R)
&\equiv& \langle [N(R) - \langle N(R) \rangle]^3\rangle  \nonumber \\
&=& \langle N^3({R})\rangle -3\langle N^2({R})\rangle \langle N({R})\rangle +2 \langle N({R})\rangle^3\nonumber \\
&=&\rho v_1(R) +3\rho^2\int_{\mathbb{R}^d}  h({\bf r}) v_2^{\mbox{\scriptsize int}}(r;R) d{\bf r}+ \rho^3\int_{\mathbb{R}^d}  \int_{\mathbb{R}^d} [g_3({\bf r}^3) -3g_2({\bf r}_{12}) +2] 
v_3^{\mbox{\scriptsize int}}({\bf r}^3;R) d{\bf r}_2 d{\bf r}_3,
\label{C3}
\end{eqnarray}
and
\begin{eqnarray}
C_4(R)
&\equiv& \langle [N(R) - \langle N(R) \rangle]^4\rangle - 3 \sigma^4(R) \nonumber \\
&=& \langle N^4({R})\rangle -4\langle N^3({R})\rangle \langle N({R})\rangle - 3 \langle N^2(R)\rangle^2 +12 \langle N^2(R)\rangle \langle N({R})\rangle^2 - 6 \langle N(R) \rangle ^4 \nonumber\\
&=&\rho v_1(R) +7\rho^2\int_{\mathbb{R}^d}  h({\bf r}) v_2^{\mbox{\scriptsize int}}(r;R) d{\bf r} + 6\rho^3\int_{\mathbb{R}^d}  \int_{\mathbb{R}^d} [g_3({\bf r}^3) -3g_2({\bf r}_{12})+2] v_3^{\mbox{\scriptsize int}}({\bf r}^3;R) d{\bf r}_2 d{\bf r}_3 \nonumber \\
&& + \,\rho^4\int_{\mathbb{R}^d}  \int_{\mathbb{R}^d}  \int_{\mathbb{R}^d}
[g_4({\bf r}^4) -4g_3({\bf r}^3) +12g_2({\bf r}_{12})  -6 ]
v_4^{\mbox{\scriptsize int}}({\bf r}^4;R)  d{\bf r}_2 d{\bf r}_3 d{\bf r}_4  - 3\left[\rho^2 \int_{\mathbb{R}^d}  g_2({\bf r})
v_2^{\mbox{\scriptsize int}}(r;R) d{\bf r}\right]^2 ,
\label{C4}
\end{eqnarray}
where we have used the following identities:
\begin{eqnarray}
v_1(R) v_2^{\mbox{\scriptsize int}}(r_{12};R)                        &=&\int_{\mathbb{R}^d} v_3^{\mbox{\scriptsize int}}({\bf r}^3;R) d{\bf r}_3,\\
v^3_1(R)                                                             &=&\int_{\mathbb{R}^d} \int_{\mathbb{R}^d}v_3^{\mbox{\scriptsize int}}({\bf r}^3;R) d{\bf r}_2 d{\bf r}_3,\\
v_1(R) v_3^{\mbox{\scriptsize int}}({\bf r}^3;R) &=&\int_{\mathbb{R}^d} v_4^{\mbox{\scriptsize int}}({\bf r}^4;R) d{\bf r}_4,\\
v^2_1(R)  v_2^{\mbox{\scriptsize int}}(r_{12};R)                     &=&\int_{\mathbb{R}^d} \int_{\mathbb{R}^d}v_4^{\mbox{\scriptsize int}}({\bf r}^4;R) d{\bf r}_3 d{\bf r}_4,\\
v^4_1(R)                                                             &=&\int_{\mathbb{R}^d} \int_{\mathbb{R}^d}\int_{\mathbb{R}^d}v_4^{\mbox{\scriptsize int}}({\bf r}^4;R) d{\bf r}_2 d{\bf r}_3 d{\bf r}_4.
\end{eqnarray}

We now derive a lower bound on $C_3(R)$ in terms of the first and second cumulants for general point processes. This easily follows
from the fact that the product $g_3({\bf r}^3) v_3^{\mbox{\scriptsize int}}({\bf r}^3;R)$ 
in the last integral of (\ref{C3}) is nonnegative for all positions and so dropping
this integral yields the following lower bound:
\begin{eqnarray}
C_3(R) &\ge& \rho v_1(R) [1-\rho v_1(R)][1-2\rho v_1(R)] +3[1-\rho v_1(R)] \left(\sigma^2(R)- \rho v_1[1-\rho v_1(R)]\right).
\label{C3-bound}
\end{eqnarray}
Similarly, dropping the positive integral in (\ref{C4}) involving the product $g_4({\bf r}^4) v_4^{\mbox{\scriptsize int}}({\bf r}^4;R)$ 
gives the following lower bound on $C_4(R)$ in terms of the
first, second and third cumulants:
\begin{eqnarray}
C_4(R) &\ge& \rho v_1(R)[1-\rho v_1(R)][2- \rho v_1(R)][3- \rho v_1(R)] -\sigma^2(R)\left(11-18 \rho v_1(R)+6 [\rho v_1(R)]^2\right)\nonumber\\
&& -3\sigma^4 +2 C_3(R)[3-2\rho v_1(R)].
\label{C4-bound}
\end{eqnarray}
These bounds are relatively tight for sufficiently small values of $R$ and can be exact (sharp)
for such $R$ for packings, as discussed in Sec. \ref{elem}.
\end{widetext}

The third- and fourth-order cumulants are directly related to the
{\it skewness} and {\it excess kurtosis}, respectively. The skewness is often
defined as
\begin{equation}
\gamma_1(R) \equiv \frac{C_3(R)}{\sigma^3(R)},
\end{equation}
which, qualitatively speaking, is a measure of the asymmetry of the probability distribution.
The excess  kurtosis,  which is a measure
of the heaviness of the ``tails" of the probability distribution, is  defined to be 
\begin{equation}
  \gamma_2(R)\equiv\frac{C_4(R)}{\sigma^4(R)}.
\end{equation}

For a general random variable, Pearson derived a lower bound on the excess kurtosis
in terms of the skewness \cite{Pe16}:
\begin{equation}
   \gamma_2(R) \ge {\gamma_1}^2(R) -2,
    \label{eq:pearson}
\end{equation}
We also apply  this lower bound  to validate our numerical results for all of our model point processes.
Such details are reported in the SM.

Both $\gamma_1$ and $\gamma_2$ are identically zero for the normal distribution
and hence such lower-order information can herald at what value of $R$
a general point configuration can be approximated by a normal distribution.
Note that the higher-order cumulants
become increasingly more complicated but can be written as a determinant
involving the moments $\langle N^m(R) \rangle$ \cite{Br04}.

In the case of a homogeneous Poisson point process, $g_n=1$ for all $n$. Hence, from the expressions
above, it immediately follows that $C_2(R)= C_3(R)=C_4(R)=\rho v_1(R)$, which are the expected well-known
results for this point process. It follows that the corresponding skewness and excess kurtosis are exactly
given by
\begin{equation}
\gamma_1(R)= \frac{1}{[\rho v_1(R)]^{1/2}} \propto R^{-d/2}
\label{gamma1-poisson}
\end{equation}
and
\begin{equation}
\gamma_2(R)= \frac{1}{\rho v_1(R)} \propto R^{-d},
\label{gamma2-poisson}
\end{equation}
respectively. Observe that both $\gamma_1(R)$ and $\gamma_2(R)$ tend to zero  in the limit $R \to \infty$,
which is consistent with the fact that the Poisson point process obeys a CLT.

\subsection{Probability Distribution Function}

It is straightforward to show that for an arbitrarily-shaped window region $\Omega$, the probability distribution function is given by \cite{Ve75,Zi77}
\begin{equation}
P[N_\Omega = M ]= \sum_{m=M}^\infty (-1)^{m-M} \frac{{\cal A}_m(\Omega)}{M! (m-M)!},
\label{P-exact}
\end{equation}
where 
\begin{align}
{\cal A}_m(\Omega) \equiv & \left\langle \frac{N_\Omega!}{(N_\Omega-m)!} \right\rangle_\Omega \nonumber \\
=&\rho^m\int_{\Omega}  \int_{\Omega} \cdots \int_{\Omega} g_m({\bf r}^m) 
d{\bf r}_1 d{\bf r}_2 \ldots d{\bf r}_m,
\end{align}
is exactly the same as the average given in (\ref{rhon}) under the assumption of homogeneity. Thus, we see that the average ${\cal A}_m(\Omega)$ is the nontrivial and common contribution to
$P[N_\Omega]$ for any specific value of $N_\Omega$.
The only differences in $P[N_\Omega]$ for different values of $N_\Omega$ are the combinatoric factors
multiplying the coefficients ${\cal A}_m(\Omega)$ in the series (\ref{P-exact}). Because these coefficients are intrinsically positive, relation (\ref{P-exact})
is an alternating series.

When $\Omega$ is a spherical window of radius $R$, it simply follows that
\begin{equation}
P[N(R)=M]= \sum_{m=M}  ^\infty (-1)^{m-M} \frac{{\cal A}_m(R)}{M! (m-M)!},
\label{P-exact-2}
\end{equation}
where 
\begin{align}
{\cal A}_m(R) \equiv & \left\langle \frac{N(R)!}{(N(R)-m)!} \right\rangle_{v_1(R)} \nonumber \\
= & \rho^m\int_{\mathbb{R}^d}  \int_{\mathbb{R}^d} \cdots \int_{\mathbb{R}^d} g_m({\bf r}^m) 
\nonumber \\
& \times  v_m^{\mbox{\scriptsize int}}({\bf r}^m;R)  d{\bf r}_2 d{\bf r}_3 \ldots d{\bf r}_m.
\label{A-m}
\end{align}
For a fixed value of $M\in\mathbb{N}_0$, $P[ N(R) < M ]$ as a function
of $R$ is a complementary cumulative distribution function, which is
associated with the ``void" probability density function $H_V(R;M)$ \cite{Tr98b}, where
$H_V(R;M)dR$ is the probability that  the distance to the $M$th nearest neighbor from an \textit{arbitrary point in
space} is between $R$ and $R+dR$.
For example, in the case $M=1$, $P[ N(R) < 1 ]=P[ N(R) = 0 ]$ is the
well-known {\it void exclusion} probability function $E_V(R)$, which is associated with the void nearest-neighbor probability density function $H_V(R)$~\cite{To90c,To02a,To10d}.
In the case of a homogeneous Poisson point process, we can immediately recover the exact result 
(\ref{P}) for the number distribution from relations (\ref{P-exact-2}) and (\ref{A-m}) using the fact that $g_n=1$ for all $n$ and the identity
\begin{equation}
v_1(R)^m=\int_{\mathbb{R}^d}  \int_{\mathbb{R}^d} \cdots \int_{\mathbb{R}^d}v_m^{\mbox{\scriptsize int}}({\bf r}^m;R)  d{\bf r}_2 d{\bf r}_3 \ldots d{\bf r}_m\;.
\end{equation}

Exact results for $P[N(R)]$ for non-Poissonian point processes are rare.
One exception is the one-dimensional model of equilibrium hard {\it rods} for which $P[N(R)]$ is known exactly \cite{Tr98b}.
In the case of  disordered equilibrium hard-disk ($d=2)$ and hard-sphere ($d=3$) packings, accurate but approximate expressions for the {\it exclusion probability} $E_V(R)\equiv P[N(R)=0]$ are available \cite{To90c,To02a}. Thus, in principle, one can extract from such formulas
approximations for ${\cal A}_m(R)$ to get corresponding approximations for $P[N(R)]$ for any $N(R) \ge 1$ for such packings.
The difficulty in ascertaining $P[N(R)]$ exactly for nontrivial models can be appreciated by appealing to 
 the {\it ghost} random sequential addition (RSA) packing process \cite{To06b}, for which the $g_m$ are known exactly for any $m$. 
While the evaluation of the integral (\ref{A-m}) for ghost RSA can be carried out exactly for very small $m$, its exact determination becomes impossible for general values of $m$.
This points to the importance of devising accurate numerical methods to determine the number distributions of packings.
For example, $P[N(R)]$ has be determined in simulations for equilibrium 
hard spheres and MRJ sphere packings in Ref.~\cite{Kl16}.

It has already been established rigorously that truncations of the alternating series  (\ref{P-exact-2}) for the special
case  $P[N(R)=0]=E_V(R)$ at an even and odd number of terms yield successive upper and lower bounds on the probability distribution $P[N(R)=0]$, respectively. 
Such bounds are consequences of an {\it inclusion-exclusion} principle associated with this alternating series. Here we 
make the simple observation that the same inclusion-exclusion principle applies for any value of $N(R)$. The first several
of such bounds are given by
\begin{align}
P[N(R)] \le& \frac{{\cal A}_{N}(R)}{N(R)!} \\
P[N(R)] \ge & \frac{{\cal A}_{N}(R)}{N(R)!} - \frac{{\cal A}_{N+1}(R)}{N(R)!} \\
P[N(R)] \le &  \frac{{\cal A}_{N}(R)}{N(R)!} - \frac{{\cal A}_{N+1}(R)}{N(R)!} +\frac{{\cal A}_{N+2}(R)}{2! N(R)!}\\
P[N(R)] \ge &  \frac{{\cal A}_{N}(R)}{N(R)!} - \frac{{\cal A}_{N+1}(R)}{N(R)!} +\frac{{\cal A}_{N+2}(R)}{ 2! N(R)! } \nonumber \\
& \quad -\frac{{\cal A}_{N+3}(R)}{3! N(R)!},
\end{align}
where ${\cal A}_N(R)$ is a shorthand for ${\cal A}_{N(R)}(R)$.
These bounds  become increasingly sharper as more terms are included.  Moreover, these bounds can be sharp (exact) for sufficiently small $R$
for sphere packings, as discussed in Sec. \ref{elem}. We utilize these bounds to validate our simulations
results for all models across dimensions in the SM.

\subsection{Elementary Results for Packings}
\label{elem}

Here we obtain some  general exact results for the first few cumulants and distribution functions for packings of identical spheres of diameter $D$, 
whether disordered or not. Results that apply to lattice packings are also derived.

Because no two spheres can overlap when their centers are separated by a  distance less than or equal to $D$, 
the cumulants can be written explicitly for $d \ge 1$ and $R \le D/2$, since such a window region
can accommodate at most a single sphere. For example, this means that integral involving $g_2$ in the last line of Eq. (\ref{C2})
is identically zero, and hence for $R \le D/2$,
\begin{equation}
 C_2(R)= \rho v_1(R) [1-\rho v_1(R)],
\end{equation}
which was noted by Torquato and Stillinger \cite{To03a}. More generally, noting that $\langle N(R)^m \rangle=\rho v_1(R)$ for all $m$ because all terms in (\ref{N3}) and (\ref{N4}) involving either
$g_2$, $g_3$ and $g_4$, we have from (\ref{C3}) and (\ref{C4}) for $d \ge 1$ that for $R \le D/2$,
\begin{equation}
 C_3(R)=  \rho v_1(R) [1-\rho v_1(R)][1-2\rho v_1(R)],
\label{C3-small}
\end{equation}
\begin{equation}
 C_4(R)= \rho v_1(R) [1-7\rho v_1(R)+12\rho^2 v^2_1(R)-6\rho^3 v^3_1(R)],
\label{C4-small}
\end{equation}
This means that for $R \le D/2$, 
\begin{equation}
\gamma_1(R)= \frac{1 -2\rho v_1(R)}{[\rho v_1(R) (1-\rho v_1(R))]^{1/2}}
\end{equation}
and
\begin{equation}
\gamma_2(R)= \frac{1-7\rho v_1(R)+12\rho^2 v^2_1(R)-6\rho^3 v^3_1(R)}{\rho v_1(R)[1-\rho v_1(R)]^2}.
\end{equation}
As in the Poisson case, both the skewness $\gamma_1(R)$ and excess kurtosis $\gamma_2(R)$ diverge to $+\infty$ in the limit $R \to 0$.
For $0 \le R \le D/2$, while $\gamma_1(R)$ is generally a monotonically decreasing function of $R$ that can 
be nonnegative, $\gamma_2(R)$ is generally a nonmonotonic function but also can be nonnegative.
The corresponding results for the distribution function $P[N(R)]$  for $R \le D/2$
follow immediately from (\ref{P-exact}) and (\ref{A-m}), namely,
\begin{eqnarray}
P[N(R)=0] &=& 1- \rho v_1(R),\\
P[N(R)=1] &=& \rho v_1(R),\\
P[N(R) \ge 2]&=&0.
\label{1}
\end{eqnarray}

\begin{widetext}
When $D/2 < R \le D/\sqrt{3}$ and $d\ge 2$, the window can accommodate at most two but not three hard spheres.
Therefore, following the same reasoning as above for $ R \le D/2$, we find from (\ref{C3}) and (\ref{C4}) that for $d \ge 2$ and  $0 \le R \le D/\sqrt{3}$ 
\begin{eqnarray}
C_3(R) &=& \rho v_1(R)  +3\rho^2\int_{\mathbb{R}^d}  h({\bf r})
v_2^{\mbox{\scriptsize int}}(r;R) d{\bf r} -3\rho^3v_1(R) \int_{\mathbb{R}^d} g_2({\bf r}_{12})  v_2^{\mbox{\scriptsize int}}(r;R) d{\bf r}
+2 \rho^3 v_1^3(R) \nonumber \\
 &=& \rho v_1(R) [1-\rho v_1(R)][1-2\rho v_1(R)] +3[1-\rho v_1(R)] \rho^2\int_{\mathbb{R}^d}  g_2({\bf r})
v_2^{\mbox{\scriptsize int}}(r;R) d{\bf r} 
 \nonumber \\
 &=& \rho v_1(R) [1-\rho v_1(R)][1-2\rho v_1(R)] +3[1-\rho v_1(R)] \left[\sigma^2(R)- \rho v_1[1-\rho v_1(R)]\right]
\label{C3-packing}
\end{eqnarray}
and
\begin{eqnarray}
C_4(R)&=& \rho v_1(R) +7\rho^2\int_{\mathbb{R}^d}  h({\bf r})
v_2^{\mbox{\scriptsize int}}(r;R) d{\bf r} 
+6\rho^3\int_{\mathbb{R}^d}  \int_{\mathbb{R}^d} [-3g_2({\bf r}_{12})+2] v_3^{\mbox{\scriptsize int}}({\bf r}^3;R) d{\bf r}_2 d{\bf r}_3 \nonumber \\
&+&\rho^4\int_{\mathbb{R}^d}  \int_{\mathbb{R}^d}  \int_{\mathbb{R}^d}
[12g_2({\bf r}_{12})   -6]
v_4^{\mbox{\scriptsize int}}({\bf r}^4;R)  d{\bf r}_2 d{\bf r}_3 d{\bf r}_4-3\left[\rho^2 \int_{\mathbb{R}^d}  g_2({\bf r})
v_2^{\mbox{\scriptsize int}}(r;R) d{\bf r}\right]^2  \nonumber \\
&=& \rho v_1(R) [1-7\rho v_1(R)+12\rho^2 v^2_1(R)-6\rho^3 v^3_1(R)] +7\rho^2\int_{\mathbb{R}^d}  g_2({\bf r})
v_2^{\mbox{\scriptsize int}}(r;R) d{\bf r}  - 18 \rho^3v_1(R) \int_{\mathbb{R}^d}   g_2({\bf r})v_2^{\mbox{\scriptsize int}}(r;R) d{\bf r} 
 \nonumber \\
&+&12 \rho^4v_1^2(R) \int_{\mathbb{R}^d} 
12g_2({\bf r}) v_2^{\mbox{\scriptsize int}}(r;R) d{\bf r}  
- 3  \left[\rho^2 \int_{\mathbb{R}^d} g_2({\bf r}) v_2^{\mbox{\scriptsize int}}(r;R) d{\bf r}  \right]^2 \nonumber\\
&=& \rho v_1(R) [1-7\rho v_1(R)+12\rho^2 v^2_1(R)-6\rho^3 v^3_1(R)]  \nonumber \\
&& + \,[7 -18\rho v_1(R)+12 \rho^2 v_1^2(R)]\, \rho^2\int_{\mathbb{R}^d}  g_2({\bf r})
v_2^{\mbox{\scriptsize int}}(r;R) d{\bf r} - 3 \left[\rho^2\int_{\mathbb{R}^d} g_2({\bf r}) v_2^{\mbox{\scriptsize int}}(r;R) d{\bf r}  \right]^2 \nonumber\\
&=& \rho v_1(R) [1-7\rho v_1(R)+12\rho^2 v^2_1(R)-6\rho^3 v^3_1(R)] +[7 -18\rho v_1(R)+12 \rho^2 v_1^2(R)]
[\sigma^2(R) -\rho v_1(R)(1-\rho v_1(R))]  \nonumber \\
&& -\, 3\, [\sigma^2(R) -\rho v_1(R)(1-\rho v_1(R))]^2.
\end{eqnarray}

\noindent We see that both $C_3(R)$ and $C_4(R)$ are given purely in terms of the mean $\langle N(R)\rangle=\rho v_1(R)$
and number variance $\sigma^2(R)$ for such $R$. Observe also that formula (\ref{C3-packing}) is identical
the lower bound (\ref{C3-bound}) for general point processes. It is seen that for $R \le D/2$, relations (\ref{C3-small}) and (\ref{C4-small}) are recovered,
as expected. Finally, for $d=1$, these formulas actually apply for $R \le D$.

When the window can accommodate at most three spheres, the fourth cumulant $C_4(R)$ for $0 \le R \le R_*(d)$ is exactly given 
in terms of the first three cumulants, i.e.,
\begin{eqnarray}
C_4(R)&=&\rho v_1(R)[1-\rho v_1(R)][2- \rho v_1(R)][3- \rho v_1(R)] -\sigma^2(R)\left[11-18 \rho v_1(R)+6 [\rho v_1(R)]^2\right]\nonumber\\
&& -3\sigma^4 +2 C_3(R)[3-2\rho v_1(R)],
\label{C4-packing}
\end{eqnarray}
where $R_*(d)>  D/\sqrt{3}$ is a threshold that depends on the space dimension $d$. For example,
$R_*(1)=3 D/2$, $R_*(2)=\sqrt{2} D/2$, and  $R_*(3)=\sqrt{3/8} D$. Note that formula (\ref{C4-packing}) is identical to
the lower bound (\ref{C4-bound}) for general point processes.

When the spherical window can accommodate at most two spheres, implying that $R \le D/\sqrt{3}$ for $d \ge 2$,
three-body and higher-order terms in the series  (\ref{A-m}) for the probability distribution 
vanish identically, yielding the following  exact result for  $P[N(R)]$:  
\begin{eqnarray}
P[N(R)=0] &=& [ 1-\frac{1}{2}\rho v_1(R) ][1-\rho v_1(R)]  +\frac{\sigma^2(R)}{2}, \\
P[N(R)=1] &=& 2\rho v_1(R)-[\rho v_1(R)]^2 -\sigma^2(R) ,\label{P1}\\
P[N(R)=2] &=& \frac{\sigma^2(R)}{2}- \frac{\rho}{2} v_1(R)[1-\rho v_1(R)] ,\label{P2}\\
P[N(R) \ge 3]&=&0.
\label{2}
\end{eqnarray}
We see that for such windows, the entire distribution
function is completely determined by the first and second cumulants.
A nontrivial upper bound on the variance $\sigma^2(R)$ of a packing follows immediately from (\ref{P1})
and the fact that $P[N(R)=1]$ must be nonnegative, i.e.,
\begin{equation}
\sigma^2(R) \le \rho v_1(R)[2-\rho v_1(R)].
\end{equation}
For the same reasons, relation (\ref{P2}) yields the following general lower bound on the variance
\begin{equation}
\sigma^2(R) \ge \rho v_1(R)[1-\rho v_1(R)].
\end{equation}
As before, these formulas actually apply for $R \le D$ for $d=1$.

Similarly, when the spherical window can accommodate at most three spheres, implying that $0 \le R \le R_*(d)$,
four-body and higher-order terms in the series  (\ref{A-m}) for $P[N(R)]$ 
vanish identically, yielding the following  exact result for  $P[N(R)]$:  
\begin{eqnarray}
    P[N(R)=0] &=& 1- \rho v_1(R) -\frac{\rho v_1(R)}{6}[1-\rho v_1(R)][5-\rho v_1(R)] + \frac{1}{2} [2-\rho v_1 (R)]\sigma^2(R) - \frac{C_3 (R)}{6}, \label{n0} \\
    P[N(R)=1] &=& \frac{1}{2} \rho v_1 (R) [2-\rho v_1(R)][3-\rho v_1(R)] \label{n1} 
+\frac{1}{2}\qty[-5 + 3 \rho v_1(R)] \sigma^2 (R) + \frac{1}{2}C_3 (R) ,\\
    P[N(R)=2] &=& -\frac{\rho v_1(R)}{2}[1-\rho v_1(R)][3-\rho v_1(R)] + \frac{1}{2}\qty[4-3\rho v_1 (R)]\sigma^2 (R) - \frac{1}{2} C_3 (R) ,\label{n2}\\
    P[N(R)=3] &=& \frac{\rho v_1(R)}{6}[1-\rho v_1(R)][2-\rho v_1(R)]
    -\frac{\sigma^2(R)}{2}[1-\rho v_1(R)] +\frac{C_3(R)}{6}, \label{n3}\\
    P[N(R)\ge 4] &=&0.
\end{eqnarray}
Thus, for such $R$, the entire probability distribution
function is completely determined by the first three cumulants.
The nonnegativities of the probabilities $P[N(R)=0]$ [Eq. (\ref{n0})] and $P[N(R)=2]$ [Eq. (\ref{n2})] yield upper bounds on $C_3(R)$ 
in terms of the first two cumulants and so the minimum of these two upper bounds are to be chosen.
Similarly, the  $P[N(R)=1]$ [Eq. (\ref{n1})] and  $P[N(R)=3]$ [Eq. (\ref{n3})] yield lower bounds on $C_3(R)$
 in terms of the first two cumulants and so the maximum of these two lower bounds are to be chosen.
In the SM, we demonstrate that  the aforementioned exact results for $C_3(R)$, $C_4(R)$ and certain $P[N(R)]$
 are in excellent agreement with our corresponding simulations data for sphere packings examined in the paper across dimensions.
\end{widetext}

More generally, for any packing of identical spheres, a spherical region of radius $R$ can accommodate
a   maximum number of spheres, denoted by  $N_{max}(R)$. This maximal number
can be determined from tabulations of the so-called {\it densest local packings}
for a finite range of particle numbers in both two \cite{Ho10b} and three \cite{Ho11a} dimensions.
Therefore, the number distribution $P[N(R)]$ for a packing is generally far from a
normal distribution for a finite-sized window, since it must have compact support such that 
it is zero for $N(R) > N_{max}(R)$, i.e.,
\begin{equation}
P[N(R)] = 0 \quad \mbox{for}\; N(R) > N_{max}(R).
\end{equation}
In such instances, the distribution function is determined by a finite set of moments, i.e., the first, second, $\ldots$, $N_{max}$th moments. 
Moreover, for any dense packing or point process in which the nearest neighbor from a particle
is narrowly distributed (e.g., ``strongly" stealthy systems described below and in Sec. \ref{stealth}), 
$P[N(R)]$ will be zero for $N(R) < N_{min}(R)$, where  the cut-off value $N_{min}(R)$ grows with $R$.
This situation prevents a strict CLT from applying for finite-sized windows.

Another important observation is that for point processes in which the ``hole" radius $R$
is bounded from above by $R_{max}$, the probability of finding a spherical window
with radius $R> R_{max}$ must be zero, i.e., $P[N(R)=0]= 0$ for $R> R_{max}$, which of
course is non-Gaussian behavior. The cut-off value $R_{max}$ for a point process in $\mathbb{R}^d$ is
its {\it covering radius} \cite{To10d}. Processes with this bounded-hole property include
periodic packings with a finite basis \cite{Zh17a}, quasicrystals \cite{Le84},
as well as  the saturated random sequential addition packing process (see Sec. \ref{rsa}). Disordered stealthy
point processes have bounded holes \cite{Zh17a,Gh18}, as discussed in Sec. \ref{stealth}.

We note that for the hypercubic lattice $\mathbb{Z}^d$ scaled by $D$ (see Sec. \ref{cubic} for precise definition),
all of the relations derived above for the skewness, excess
kurtosis and distribution function for the situation $R \le D/\sqrt{3}$ actually apply as well for the larger range $R \le D/\sqrt{2}$
when $d \ge 2$, where $D$ is the lattice spacing. In fact, in the case of the scaled integer lattice $\rho^{-1}\mathbb{Z}$ at number density $\rho$,
we can obtain an exact formula for $P[N(R)]$ by invoking the key idea of Ref. \cite{To03a} to yield  the exact result for the local number variance,
namely, the number of points inside a window of radius $R$ can only take two values, either $N_R$ or $N_R +1$, where $N_R\equiv\lfloor{\rho 2R}\rfloor$ and $\lfloor{x}\rfloor$ is the floor function of a real number $x$. The probability distribution for all $N(R)$ is given by
\begin{eqnarray}
P[N(R)< N_R] &=& 0, \\
P[N(R)=N_R] &=& 1- \{\rho 2R\} ,\\
P[N(R)=N_R+1] &=&  \{\rho 2R\},\\
P[N(R) > N_R+1]&=&0,
\label{3}
\end{eqnarray}
where $\{x\}\equiv x-\lfloor{x}\rfloor$ is the fractional part of a positive number $x$.
Thus, this skewed distribution is 
highly non-Gaussian with nonexistent left or right tails for almost all values of $N(R)$, implying values of the skewness
and excess kurtosis that are generally far from zero for almost $R$. From the distribution function (\ref{3}) and relation (\ref{mom-P}),
we can immediately obtain the first several cumulants:
\begin{eqnarray}
\sigma^2(R)&=&\{\rho 2R\} (1-\{\rho 2R\} ),\label{sig-int}\\
\gamma_1(R)&=&\frac{1-2\{\rho 2R\}}{[\{\rho 2R\} (1-\{\rho 2R\}) ]^{1/2}},\\
\gamma_2(R)&=&\frac{1-6\{\rho 2R\}(1-\{\rho 2R\})}{\{\rho 2R\} (1-\{\rho 2R\})},
\end{eqnarray}
which are all periodic functions with period $\rho 2R$.
Relation (\ref{sig-int}) for the variance was given in Ref. \cite{To03a}.
The reader is referred to the top panel of Fig. \ref{cumulants}, which shows plots of the variance, skewness and excess kurtosis
as a function of $R$ for $\mathbb{Z}$. The highly discrete nature of the number distribution for the integer
lattice extends to that for the hypercubic lattice $\mathbb{Z}^d$ for $d \ge 2$,
as will see in Sec. \ref{results}.

\section{ Nonhyperuniform and  Hyperuniform Models}
\label{models}

We consider eight different models of statistically homogeneous point processes
in two and three dimensions:  five nonhyperuniform models, one of which is
anti-hyperuniform (hyperplanes intersection process or HIP), and three hyperuniform models. Analogous
models are also examined in one dimension, except for HIP, which is not defined in this dimension.
The reader is referred to Figs. \ref{models-nonhyper} and \ref{models-hyper} for representative images of configurations
for each of the models in two dimensions.

It is useful to recall scaling relations for hyperuniform and nonhyperuniform point processes.
Consider any homogeneous point process in $\mathbb{R}^d$ for which
the structure factor has the following power-law behavior as the wavenumber
tends to zero:
\begin{equation}
S({\bf k}) \sim |{\bf k}|^{\alpha} \qquad (|{\bf k}| \to 0).
\end{equation}
This scaling implies that the total correlation function $h(\bf r)$ has the corresponding
power-law behavior $1/|{\bf r}|^{d+\alpha}$ for large $|\bf r|$ \cite{To18a}. 
For hyperuniform systems, the exponent $\alpha$ is a positive constant, which
implies that there are three different scaling regimes (classes)
that describe the associated large-$R$ of the number variance \cite{To03a,Za09,To18a}:
\begin{align}  
\sigma^2(R) \sim 
\begin{cases}
R^{d-1}, \quad\quad\quad \alpha >1 \qquad &\text{(Class I)}\\
R^{d-1} \ln R, \quad \alpha = 1 \qquad &\text{(Class II)},\\
R^{d-\alpha}, \quad 0 < \alpha < 1\qquad  &\text{(Class III)}.
\end{cases}
\label{sigma-hyper}
\end{align}
By contrast, for any nonhyperuniform system, it follows from
the asymptotic analysis given in Ref. \cite{To18a} that
\begin{align}  
\sigma^2(R) \sim 
\begin{cases}
R^{d}, & \alpha =0 \quad \text{(typical nonhyperuniform)}\\
R^{d-\alpha}, & \alpha < 0 \quad \text{(anti-hyperuniform)}.\\
\end{cases}
\label{sigma-nonhyper}
\end{align}
The scaling for the anti-hyperuniform instance can be obtained using an asymptotic
analysis of either the direct-space representation (\ref{local1})
or the Fourier-space representation (\ref{local2}) of the number variance,
as derived in Ref.~\cite{To20b}.
The typical nonhyperuniform scaling in (\ref{sigma-nonhyper})
results from the fact that $S(0)$ is bounded and, indeed, the implied
constant multiplying $R^d$ is proportional to $S(0)$.

Any nonhyperuniform point process for which $S(0) >1$ has a large-$R$ asymptotic number variance $\sigma^2(R)$ that is larger
than that for a Poisson point process [$S(0)=1$] with the same mean $\langle N(R)\rangle$.
We call such a nonhyperuniform point process {\it super-Poissonian}.  Two examples of super-Poissonian
point processes studied in this work are the Poisson cluster and HIP point processes described below.

\subsection{Nonhyperuniform Processes}
\label{nonhyper}

\subsubsection{Poisson Point Process}
\label{poisson}

	A homogeneous Poisson point process in $\mathbb{R}^d$  has a structure factor 
$S(k)=1$ for all $k$ and hence is nonhyperuniform. At unit mean density ($\rho=1$) this process is generated within a hypercubic simulation box of fixed volume $V$ under periodic boundary conditions
by a two-step procedure. First,  we choose a random number $N$ from the Poisson distribution (\ref{P}) with
intensity or mean $\rho V=V$ and then place $N$ points in the simulation box uniformly.

\subsubsection{Equilibrium Packings}

We also consider equilibrium packings of identical sphere (Gibbs hard-sphere processes)
across the first three space dimensions. For $d=2$ and $d=3$, we examine disordered states that lie
along the stable liquid branch \cite{Ha86,To02a} as well as disordered states in one dimension, all of which are nonhyperuniform.
We generate such equilibrium packings using the well-established Metropolis numerical scheme \cite{Ha86,To02a}.
All configurations that we generate are well away from jamming points and hence all are nonhyperuniform with 
bounded $S(0)$ (see Table \ref{parameters}).

\subsubsection{Random Sequential Addition Packings}
\label{rsa}

The random sequential addition (RSA) process is a time-dependent (nonequilibrium) procedure 
that generates disordered sphere packings in $\mathbb{R}^d$ \cite{Re63,Wi66,Fe80,Co88,To06a,Zh13b}. 
Starting with an empty but large volume in  $\mathbb{R}^d$, the RSA process is produced by randomly, irreversibly, and sequentially placing nonoverlapping 
spheres into the volume. This procedure is repeated for ever increasing volumes and then
an appropriate infinite-volume limit is obtained.
In practice,  hard spheres are randomly
and sequentially placed into a large fundamental cell under periodic boundary conditions
and subject to a nonoverlap
constraint: If a new sphere does not overlap with any existing
spheres, it will be added to the configuration; otherwise, the
attempt is discarded. One can stop the addition process at any
time $t$, obtaining RSA configurations with a range of packing fractions
$\phi(t)$ up to the maximal saturation value $\phi_s \equiv \phi(\infty)$, which imposes a {\it bounded-hole}  property \cite{Zh17a}.
For identical spheres, which we consider here, $\phi_s \approx 0.74, 0.55,$ and $0.38$
for $d=1,2,$ and $3$, respectively \cite{Re63,Fe80,Co88,To06d,Zh13b}.
The pair correlation function $g_2(r)$ is known exactly only in one dimension
\cite{Bo94}. The structure factors at the saturation states across dimensions have been determined
numerically \cite{To06d,Zh13b}. These results reveal that saturated RSA packings
are nonhyperuniform, even if the values of $S(0)$ are relatively small (see Table \ref{parameters}).

\subsubsection{Poisson Cluster Process}

The \textit{Poisson cluster process} is an example of a strongly 
clustering point process with large density fluctuations on large 
length scales, i.e., with a large but finite value of $S(0)$,
and hence is a nonhyperuniform system that is far from being hyperuniform.
The construction of the cluster process starts from a homogeneous 
Poisson point process of intensity  $\rho_p$~\cite{La17}.
Each point of the Poisson point process is the center of a cluster of points.
The number of points in each cluster is independent and follows a 
Poisson distribution with mean value $c$.
In our specific model, the positions of the points relative to the 
center of the cluster follows an isotropic Gaussian distribution with 
standard deviation  $r_0$,
which can be regarded to be the characteristic length scale of a single cluster.
This model is also known as a (modified) Thomas point process, which is 
an example of a Neyman-Scott process~\cite{Chi13, Il08}.
In the infinite-volume limit, the pair correlation function in $\mathbb{R}^d$ is exactly given by~\cite{Il08}:
\[
g_2(r) = 1 + \frac{c}{\rho(4\pi r_0^2)^{d/2}}e^{-\frac{r^2}{4 r_0^2}}.
\]
Thus, the corresponding structure factor  for any $d$ is given by
\begin{equation}
S(k)= 1+ c e^{- k^2 r_0^2}.
\end{equation}
and hence such processes are nonhyperuniform and super-Poissonian with $S(0)=1+c$.
 To simulate the process, which is straightforward, we use 
periodic boundary conditions and the following parameters across the first three space dimensions:
$r_0=1$ and unit number density $\rho=\rho_pc=1$ such that  $\rho_p=0.1$ and $c=10$. 
For such parameters, $S(0)=11$ across dimensions (see Table \ref{parameters}).

\subsubsection{Hyperplanes Intersection Process}\label{sec:HIP}

The \textit{hyperplanes intersection process} (HIP) is 
hyperfluctuating~\cite{To18a}, i.e., its number 
variances scales faster than the volume of the observation 
window and $\lim_{k\rightarrow 0} 
S(k)=\infty$~\cite{He06, Kl19a}.
This antihyperuniform and super-Poissonian point process is defined as the vertices (i.e., intersections) of a 
Poisson hyperplane process, that is, of randomly and independently 
distributed hyperplanes~\cite{Schn08,  Chi13}.
 In the infinite-volume limit, the pair correlation function in $\mathbb{R}^d$ for any $d\ge 2$ is exactly given by~\cite{He06}:
\[
g_2(r) = 1 + \sum_{k=1}^{d-1}\binom{d-1}{k}\left(\frac{\omega_{d-k}}{\omega_{d}}\right)^2\left(\frac{d\omega_d}{\omega_{d-1}}\right)^k\frac{1}{( s r)^k},
\]
where $s$ is the specific surface of the hyperplane and $\omega_d$ denotes the volume of a $d$-dimensional sphere
of unit radius. The number density $\rho$ is determined by the specific surface area $s$ 
of the Poisson hyperplane process (which is the only parameter of the 
isotropic HIP):
\begin{align}
  \rho = \omega_d \left(\frac{\omega_{d-1}}{d\omega_d}\right)^d s^d.
\end{align}
According to (\ref{sigma-nonhyper}), because $\alpha=1$ for any $d$, the number variance has the large-$R$ scaling $\sigma^2(R) \sim R^{2d-1}$.
Clearly, this process does not exist for $d=1$.
To simulate this process, we cannot employ periodic boundary conditions; rather, we 
circumscribe the cubic simulation box by a hypersphere and then generate 
intersecting hyperplanes that are Poisson distributed ~\cite{Kl19a}.
The orientation of the hyperplanes is uniformly distributed on the 
unit sphere and the distance of the hyperplanes  to the center of the 
simulation box is uniformly distributed between zero and the radius of 
the circumsphere. The point process at unit number density is then simulated by computing 
all intersections of hyperplanes (within the  circumsphere).

\subsection{Hyperuniform Processes}

\subsubsection{Hypercubic Lattice}
\label{cubic}

Interestingly, the problem of determining number fluctuations in lattices
has deep connections to number theory, including Gauss's circle problem \cite{Ga31} and its
generalizations \cite{To18a} as well as the Epstein zeta function \cite{Sa06}, which is directly 
related to the minimization of the number variance \cite{To03a,Za09,To18a}.
All periodic point patterns in $\mathbb{R}^d$, including Bravais lattices, are hyperuniform 
of class I \cite{To03a,Za09,To18a}.
The hyperuniformity concept enables one to rank order lattices and other periodic point patterns
according to the degree to which they suppress large-scale density fluctuations as defined
by the number variance \cite{To03a,Za09,To18a}.

For the purposes of this investigation, it is sufficient to consider the higher-order
fluctuations of the  {\it hypercubic} lattice $\mathbb{Z}^d$ is defined by
\begin{equation}
\mathbb{Z}^d=\{(x_1,\ldots,x_d): x_i \in {\mathbb{ Z}}\} \quad \mbox{for}\; d\ge 1
\end{equation}
where $\mathbb{Z}$ is the set of integers ($\ldots -3,-2,-1,0,1,2,3\ldots$)
and $x_1,\ldots,x_d$ denote the components of a lattice vector.

\subsubsection{Uniformly Randomized Lattice} 

It is well known that if the sites of a lattice are stochastically displaced by  certain finite distances, the scattering intensity (structure factor)
 inherits the Bragg peaks (long-range order) of the original lattice,
in addition to a diffuse contribution.
It has recently been  demonstrated that these Bragg peaks can be hidden in the 
scattering pattern for certain independent and identically distributed
perturbations. We have referred to this protocol as the
\textit{uniformly randomized lattice} (URL) model \cite{Kl20a}.
The underlying long-range order can be ``cloaked",  under certain conditions, in the sense that it
cannot be reconstructed from the pair-correlation function alone.
Here we generate the URL model using  the hypercubic lattice $\mathbb{Z}^d$ 
and displace each lattice
point by a random vector that is uniformly distributed in a rescaled fundamental cell of the
lattice with $a F\equiv[-a/2,a/2)^d$. The constant $a$ controls the strength of perturbations.
Counterintuitively, the long-range order suddenly disappears at certain
discrete values of $a$ and reemerges for stronger perturbations, as we
will show. Here we cloak the Bragg peaks of $\mathbb{Z}^d$ using the special value $a=1$.
Such cloaked URLs are hyperuniform such that $S(k) \sim k^2$ in the limit $k \to 0$
and hence are of class I [see Eq. (\ref{sigma-hyper})].

\subsubsection{Stealthy Hyperuniform Process}
\label{stealth}

Stealthy hyperuniform processes are defined by a structure factor that vanishes in a spherical region
around the origin, i.e., $S({\bf k})=0$ for $0<|{\bf k}|\leq K$. Such point processes
are hyperuniform of class I; see Eq. (\ref{sigma-hyper}).
A powerful procedure that enables one to generate high-fidelity stealthy
hyperuniform point patterns  is the  {\it collective-coordinate} optimization technique
\cite{Fa91,Uc04b,Uc06b,Ba08,To15,Zh15a}. This optimization methodology involves
 finding the  highly degenerate ground states of a
class of bounded pair potentials with compact support in Fourier space, which
 are stealthy and  hyperuniform by construction. The {\it control parameter} $\chi$
is a dimensionless measure of the ratio of
constrained degrees of freedom (i.e., wave vectors contained within the cut-off
wavenumber $K$) to the total degrees of freedom (approximately $d N$) in such an
optimization procedure.  A point configuration with a small
value of $\chi$ (relatively unconstrained) is disordered, and as $\chi$ increases, the short-range
order increases within a disordered regime ($ \chi < 1/2$ for $d=2$ and $d=3$) \cite{To15}. For $d=1$,
stealthy hyperuniform states can be disordered for $\chi < 1/3$ \cite{Zh15a}.
Here we use the collective-coordinate procedure to generate ``entropically-favored" disordered stealthy point
processes by first performing molecular dynamics simulations
at sufficiently low temperatures and then  minimizing the energy to obtain ground states
with exquisite accuracy \cite{Zh15a}. Importantly, stealthy states possess the  bounded-hole property \cite{Zh17a,Gh18}
and hence, as discussed in Sec. \ref{elem}, $P[N(R)=0]= 0$ for $R> R_{max}(\chi)$, where $R_{max}(\chi)$ is the
radius of the largest hole in space dimension $d$, which depends on the control parameter $\chi$.

\section{Gaussian Distance Metric}
\label{distance}

As noted in the Introduction, we are interested in ascertaining how
large $R$ must be such that $P[N(R)]$ is well approximated by the
normal distribution. There are several candidate ``distance'' metrics that we have considered
that could be used to quantify proximity to the Gaussian distribution
(beyond the skewness and excess kurtosis).
One possible distance metric that we considered is the \textit{Kolmogorov-Smirnov} test
statistic~\cite{He02}. While it is statistically robust, we found it be too insensitive for our
purposes. The \textit{Kullback-Leibler} divergence (also called the {\it relative
entropy}) is a well-known measure of the difference between probability
distributions~\cite{Ku51}.
However, it is not well-defined for comparing a discrete to a continuous
distribution; see the SM for details.

A recent study that considered distance metrics between a pair of general functions
that depend on $d$-dimensional vectors was based on the integrated squared
difference in $\mathbb{R}^d$, i.e., an $L_2$ distance metric~\cite{Wa20a}.
These authors also found that the \textit{Kullback-Leibler} distance was
not useful for their purposes.
These findings motivated us to consider distance metrics based
on the squared difference between the Gaussian and number distributions.

We identify here  two different contributions to the ``distance''
between a number distribution $P[N(R)]$ and a Gaussian distribution:
(1) deviations in the functional form of $P[N(R)]$ and
(2) the discreteness of $N(R)$ that can only approximate the
continuous Gaussian distribution.
We found that the second contribution is essentially determined by the
value of the number variance $\sigma^2(R)$, i.e., the contribution
is smaller for larger values of $\sigma^2(R)$ for the following
reason. Consider the standardized random variable $[N(R)-\langle
N(R)\rangle]/\sigma(R)$ whose discrete probability mass function is to
approximate the continuous normal density.
Then, the bin width of the probability mass function is given by
$1/\sigma(R)$ and converges to zero for $\sigma(R)\to\infty$.

Moreover, we found that the weight of the two contributions (relative to
each other) strongly depends on the representation of the number
distribution, e.g., via the characteristic function (Fourier
representation)~\cite{He02} or via direct space representations either
in discrete or continuous forms.
In fact, the choice of representation can virtually reverse the order of
the distance metrics for our point processes.
For example, a strong contribution (2) may result in a lower distance
metric for the highly skew distribution of the HIP than for the Poisson
process.
Because contribution (2) of the discreteness of $P[N(R)]$ is already
essentially given  by $\sigma^2(R)$, we here choose a representation
that focuses on deviations in the functional form of $P[N(R)]$ from that
of a Gaussian random variable.

Therefore, we define an integer-valued random variable $G(R)$, whose
probability mass function $P_{G(R)}$ is proportional to a Gaussian
distribution with the same first and second moment as our number
distribution at radius $R$. We introduce   a type of $L_2$ distance metric,   denoted by $l_2(R)$, which
is a Gaussian distance metric that employs the cumulative distribution function:
\begin{equation}
  l_2(R) \equiv \left[ \frac{1}{\sigma(R)}\sum_{n=0}^\infty |F_G(n) - F_N(n)|^2 \right]^{1/2}
\label{l2}
\end{equation}
where $F_G(n)$ is the cumulative distribution function of $G(R)$, i.e.,
\begin{equation}
 F_G(n) = \sum_{m=0}^{n} P_{G(R)}[G(R) = m],
\end{equation}
and $F_N(n)$ is the cumulative probability distribution of $N(R)$, i.e.,
\begin{equation}
 F_N(n) = \sum_{m=0}^{n} P[N(R) = m].
\end{equation}
Note that the series of the squared differences in (\ref{l2}) is scaled by $1/\sigma(R)$
because for a Gaussian distribution the range of values for which $F_G\in[\varepsilon,1-\varepsilon]$ is proportional to $\sigma(R)$.
If a particular number distribution converges (sufficiently fast) to a Gaussian distribution, $l_2(R)$ will tend to zero.
In summary, the Gaussian distance metric (\ref{l2}) is designed to be sensitive to small deviations
in the distribution functions and at the same time robust against
statistical fluctuations.

\begingroup
\squeezetable
\begin{table*}[t]
  \centering
  \caption{Simulation parameters all of the model  point processes across the first
three space dimensions  considered in the
  work. Here, $\cal{N}$ is the average of point number inside a
  fundamental cell, $N_c$ is the number of point patterns considered,
  $N_\text{window}$ is the number of observation windows per point
  pattern, and $R_{\max}$ is the largest radius of an observation window. We have also indicated
 the values of the structure factors at the origin, $S(0)$. In the cases of the equilibrium packings,
they are obtained  from  the exact result for hard rods ($d=1$) \cite{Ze27} and highly accurate approximations
for the  conditional nearest-neighbor function $G_P(r)$ in the limit $r \to \infty$ \cite{To95a, S0}.
Note that our computer simulation results are in excellent agreement with these analytical estimates
of $S(0)$ for equilibrium packings.}
  \begin{tabular}{@{\hskip 0.2in}l@{\hskip 0.2in} c c c c c c}
    \toprule
    \multicolumn{2}{c}{Models} & $\cal{N}$ & $N_c$ & $N_\text{window}$ &  $R_{\max}\rho^{1/d}$ & $S(0)$ \\
    \colrule
    \multirow{3}{*}{Antihyperuniform}
    & 2D HIP                               & $\infty$          & $10^6$           & 1      & 50  & $\infty$                \\
    & 3D HIP                               & $\infty$          & $10^6$           & 1      & 30.1  & $\infty$                \\
 & 4D HIP                               & $\infty$          & $10^4$           & 1      & 5.4  & $\infty$                \\
    \colrule
    \multirow{11}{*}{Nonhyperuniform}
    & 1D Poisson cluster                   & $105$             & $10^6$           & 1      & 50  & 10                      \\
    & 2D Poisson cluster                   & {\color{black}$205^2$}~           & {\color{black}$10^7$}~           & 1      & {\color{black}100}~  & 10                      \\
    & 3D Poisson cluster                   & $45^3$           & $10^7$           & 1      & 20  & 10                      \\
    \cline{2-7}
    & 1D Poisson \cite{comm-poisson}          & {\color{black}$\infty$}~            & {\color{black}1}~              & {\color{black}$\infty$}~ & 50  & 1                       \\
    & 2D Poisson \cite{comm-poisson}          & {\color{black}$\infty$}~            & {\color{black}1}~              & {\color{black}$\infty$}~ & {\color{black}50}~  & 1                       \\
    & 3D Poisson \cite{comm-poisson}          & {\color{black}$\infty$}~            & {\color{black}1}~              & {\color{black}$\infty$}~ & {\color{black}50}~  & 1                       \\
    \cline{2-7}
    & 1D RSA ($\phi=0.74$)                 & {\color{black}$10^7$}~            & {\color{black}$9987$}~         & $10^2$ & {\color{black}50}~  & 0.051~\cite{To06d}      \\
    & 2D RSA ($\phi=0.55$)                 & $10^4$            & $10^4$           & $10^3$ & 25  & 0.05869(4)~\cite{Zh13b} \\
    & 3D RSA ($\phi=0.38$)                 & $10^6$            & 250              & $10^4$ & 25  & 0.05581(5)~\cite{Zh13b} \\
    \cline{2-7}
    & 1D Equil.~hard rods    ($\phi=0.75$) & $5\times 10^3$      & $10^3$          & $10$ & 100 & 0.0629(2)               \\
    & 2D Equil.~hard disks   ($\phi=0.65$) & $10^4$            & $10^3$           & $10$ & 15 & 0.0260(4)               \\
    & 3D Equil.~hard spheres ($\phi=0.48$) & $10^4$            & 100              & $10^3$ & 20  & 0.022(1) \\
    \colrule
    \multirow{8}{*}{Hyperuniform}
    & 1D Cloaked URL                       & $10^4$            & $10^6$           & 1      & 50  & 0                       \\
    & 2D Cloaked URL                       & $10^4$            & $10^7$           & 1      & {\color{black}50}~  & 0                       \\
    & 3D Cloaked URL                       & $44^3$            & $10^7$           & 1      & 20  & 0                       \\
    \cline{2-7}
    & 1D Stealthy ($\chi=0.30$)            & $10^3$            & 900              & $10^3$ & 50  & 0                       \\
    & 2D Stealthy ($\chi=0.49$)            & $10^4$            & {\color{black}700}~& $10^2$ & {\color{black}50}~  & 0                       \\
    & 3D Stealthy ($\chi=0.49$)            & $10^3$            & $5.3\times 10^3$ & $10^3$ & 5   & 0                       \\
    \cline{2-7}
    & Integer lattice \cite{comm-integer}     &       {\color{black}$\infty$}~            & {\color{black}1}~              & {\color{black}$\infty$}~ & {\color{black}50}~  & 0                       \\
    & Square lattice                       & $10^4$            & 1                & $10^5$ & 50  & 0                       \\
    & SC lattice                           & $5.12\times 10^5$ & 1                & $10^4$ & 40  & 0                       \\
    \botrule
  \end{tabular}
\label{parameters}
\end{table*}
\endgroup

\section{Sampling of Moments and Number Distribution}
\label{sec:sampling}

We sample number fluctuations within a spherical window of radius $R$
for all models using a two-step procedure.
First, we randomly place the observation window in the sample (using a
uniform distribution for its center).
Second, we determine the number of points $N(R)$ within the observation
window using periodic boundary conditions, except for the  hyperplanes intersection process (HIP), as described in Sec. \ref{sec:HIP}.
To reduce computational resources, we use the same centers for all radii
that we consider.
Relevant simulation parameters and properties for each of the models
described above across dimensions are listed in Table \ref{parameters}.

Thus, we empirically determine the probability mass function $P[N(R)]$ 
and compute the mean value, variance, skewness, excess kurtosis.
We determine the first four central moments using the unbiased 
estimators from Ref.~\cite{Kl09}.
Finally, we compute the $l_2$ distance metric, as described above.
A source of systematic errors, which must be avoided for 
both the moments and distance measures, can arise when the number of 
observation windows $N_\text{window}$ per sample is too large.
This effect can cause the distance metric $l_2(R)$ to artificially 
increase again for large radii.
Therefore, we have used between 1 and $10^4$ observation windows per 
sample depending on the system size and computational cost, so that the 
the systematic error is either avoided completely or smaller than the 
effects caused by statistical fluctuations.
As a rule of thumb, we chose
$N_\text{window}$ such that the volume fraction of the union of all observation windows of the largest window radius $R_\text{max}$ in one sample should not exceed 50\%, i.e.,
$ 1-\exp\left[-N_\text{window} v_1 (R_\text{max})/V\right] < 0.5,$
where $V$ is the volume of a single sample.
Our estimators for $l_2(R)$ are statistically robust for values that are 
larger than the inverse of the total number $n$ of observation windows, 
where $n=N_\text{window}\times N_c$  with $N_c$ being the number of 
configurations.
For a finite number of samples, we empirically find that   
$l_2 (R) $ typically cannot be smaller than approximately $O(1/n)$.
Therefore, we apply a datacut and only consider radii up to
\[
  R_{\text{cut}} \equiv \min_{R>3}\{R:l_2(R)<1/\sqrt{n}\}.
\]
We apply the same datacut to the skewness and excess 
kurtosis~\footnote{We expect this datacut to be conservative, because
the $l_2$ distance metric contains information of all moments
and higher moments less numerically robustly.
By visual inspection, we found similar range of radii
for which we have reliable estimates for
$\gamma_1$, $\gamma_2$, and $l_2$.}.
We show all data without the datacut in the SM.
All simulated data are available at a
Zenodo repository \footnote{The data will be published together with the paper.}.

Another obvious source of systematic errors can arise when the size of the sampling window is not much smaller than the size of the simulation box with a fixed number of particles, i.e., when  canonical ensembles 
are employed.  It is well-known that such finite-size effects lead to an 
underestimation of the local  number variance  $\sigma^2_\text{finite}(R)$
when compared to its value in the thermodynamic limit. An empirical formula
to estimate the first-order correction to the thermodynamic
limit \cite{Ro99b} shows that the error term is proportional $S(0)$ (because 
there are larger fluctuations in the number of points per simulation 
box). Importantly, this implies that hyperuniform models, defined by  $\lim_{k\to 0}S(k)=0$, are more robust against such
finite-size effects.

\section{Results}
\label{results}

We describe results that we have obtained for the second, third and fourth cumulants,
$\sigma_2(R)$, $\gamma_1(R)$ and $\gamma_2(R)$,  as a function of  the window radius $R$
for all models across the first three space dimensions at unit number density
($\rho=1$), as well as the corresponding full probability distributions $P[N(R)]$ and the
Gaussian distance metric $l_2(R)$.
A testament to the high-precision of the data is the excellent
agreement with rigorous bounds for these quantities for general cases as well as with exact
results for the cases of packings for certain $R$ reported in Sec.~\ref{general}
(see the SM for details). 

\subsection{Cumulants $\sigma^2(R)$, $\gamma_1(R)$ and $\gamma_2(R)$ for the Models Across Dimensions}
\label{cum}

\begin{figure*}[t]
	\includegraphics[width = 1.0\textwidth]{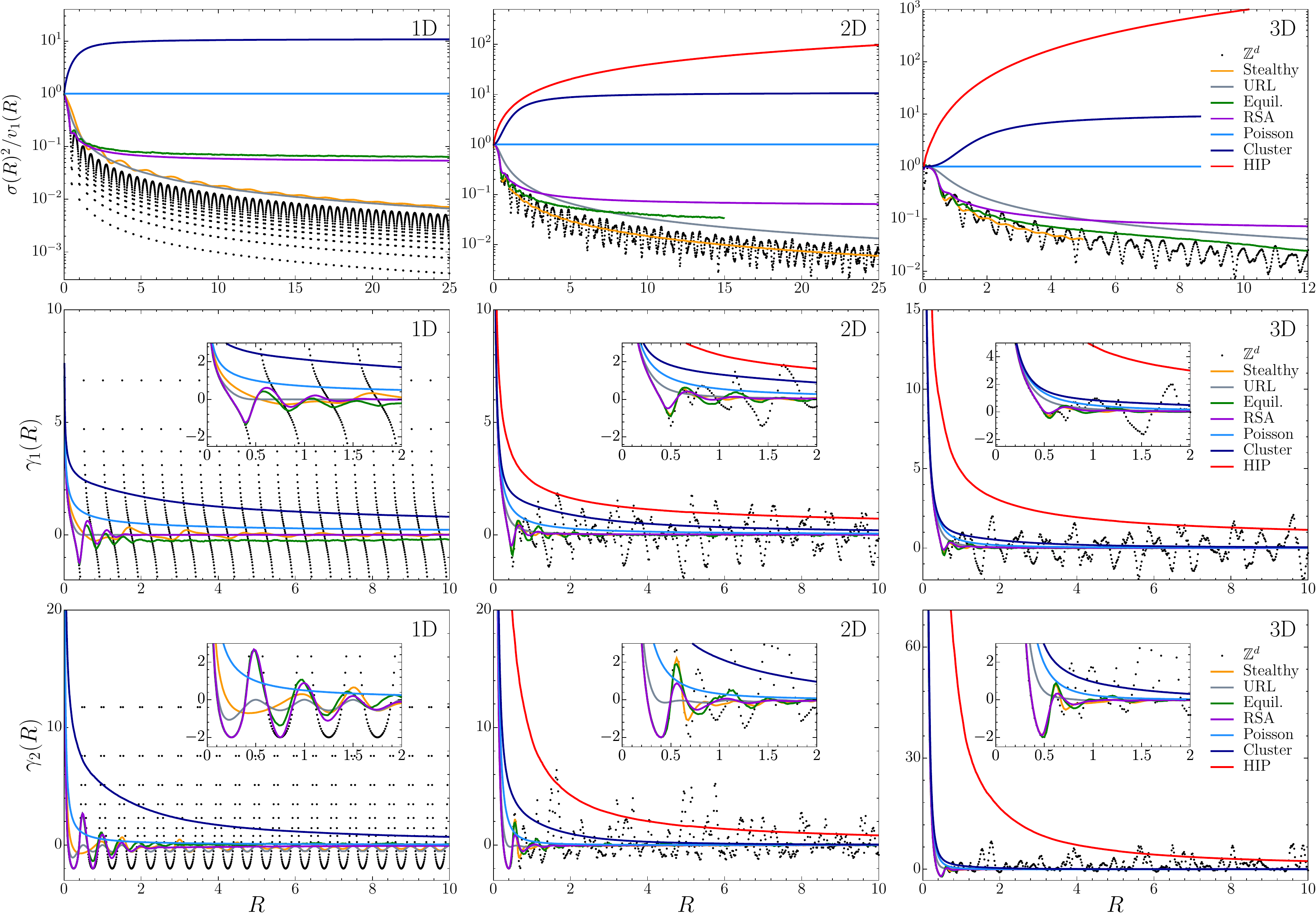}
	\caption{Graphs of the number variance, skewness and excess kurtosis versus the window radius $R$ for all considered models across the first three space dimensions.}
	\label{cumulants}
	\end{figure*}

Figure ~\ref{cumulants} shows the second, third and fourth  cumulants as a function of $R$ versus the window radius $R$ for all models for $d=1$, $d=2$ and $d=3$.
The large-$R$ asymptotic behavior of the variance is determined by the structure factor at the origin, $S(0)$ \cite{To03a}.
As expected, the scaled number variance $\sigma^2(R)/v_1(R)$ grows with $R$ for the HIP process for two-dimensional (2D)
and three-dimensional (3D) cases. For all other
nonhyperuniform models, $\sigma^2(R)/v_1(R)$ asymptotes to a constant for large $R$ in all dimensions. Of course, this scaled
variance decreases with $R$ for the three hyperuniform models
(hypercubic lattice, URL and stealthy systems).

For $d\geq 2$, we  have found via analyses given in the SM and immediately below 
that the skewness $\gamma_1(R)$ and excess kurtosis $\gamma_2(R)$  for our disordered hyperuniform systems vanish faster with increasing $R$ than 
those for nonhyperuniform systems. Among all models studied, the quantities $\gamma_1(R)$ and  $\gamma_2(R)$ vanish slowest
for the antihyperuniform HIP models for $d\ge 2$. The specific decay rates
for all models are described below.


It is noteworthy that one-dimensional systems can present fluctuation 
anomalies not present in higher dimensions.
For example, in the case of the integer lattice, the 
random variable $N(R)$ can only take at most two values for any $R$, 
which of course is abnormally non-Gaussian.
While the hypercubic lattice for $d\ge 2$ never achieves a CLT (as discussed below), the 
variance is considerably broader than that for $d=1$.
Another anomalous category is class I hyperuniform systems, which have 
bounded variance for $d=1$ [see Eq.
(\ref{sigma-hyper})] and thus any such hyperuniform point process cannot 
obey a CLT because the standardized distribution cannot converge to a 
continuous distribution. This is clearly borne out by the distance metric plots, shown in Fig.
\ref{metrics} and  Fig. S8 of the SM, for both the 1D disordered 
stealthy point process and integer lattice.

\begin{figure*}[t]
	\centering{\includegraphics[width = 0.48\textwidth]{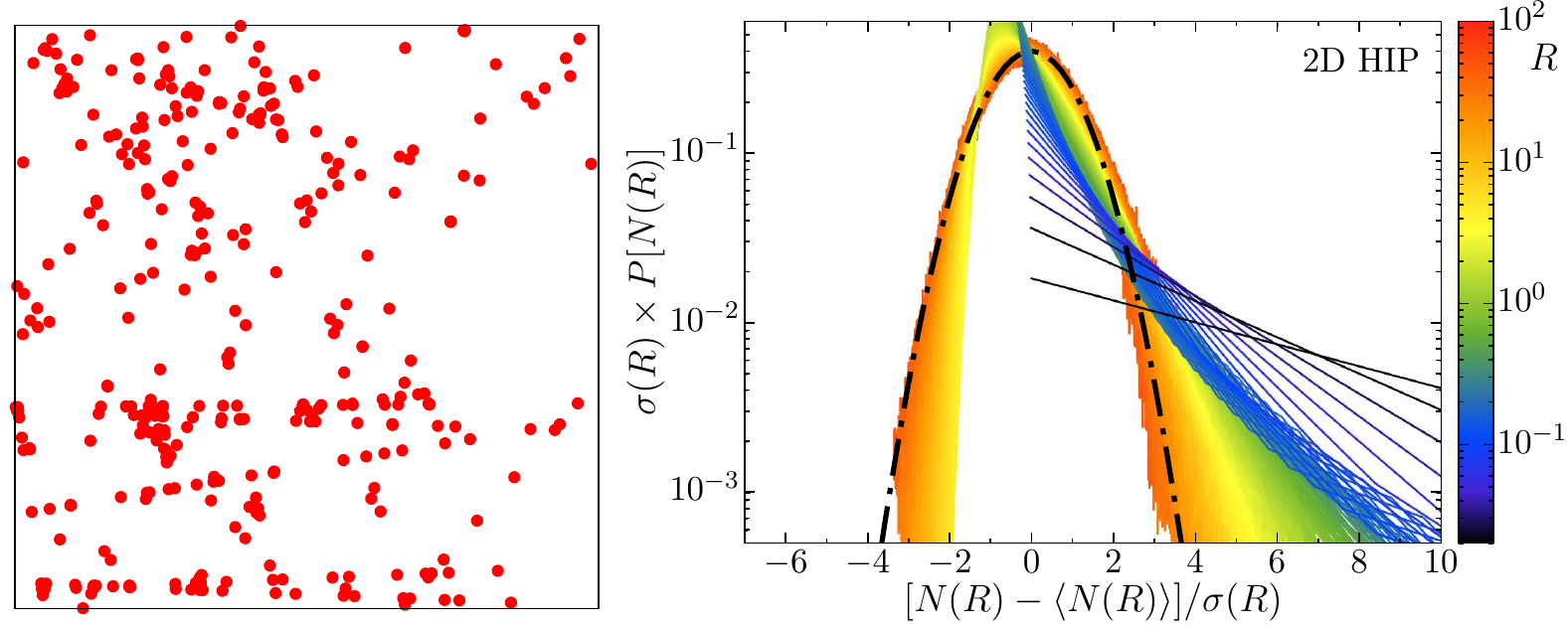}}\hfill%
	\centering{\includegraphics[width = 0.48\textwidth]{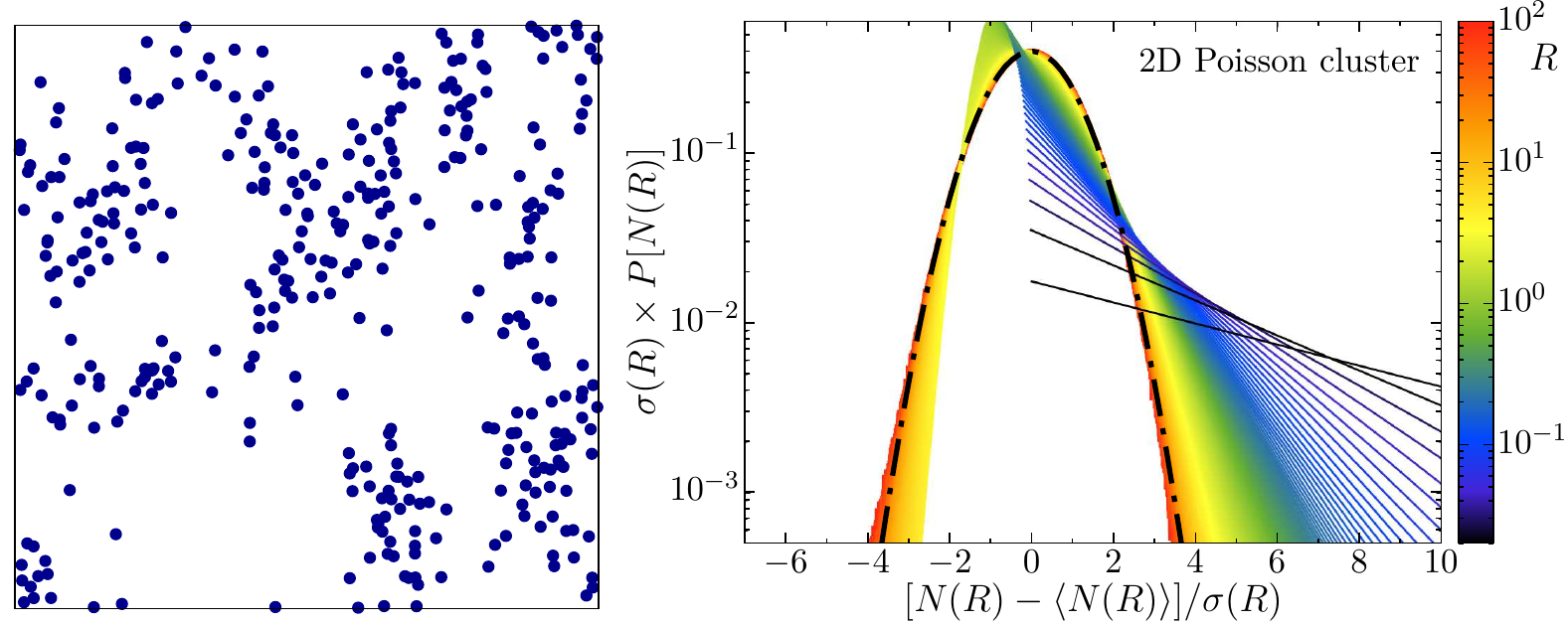}}\\
	\centering{\includegraphics[width = 0.48\textwidth]{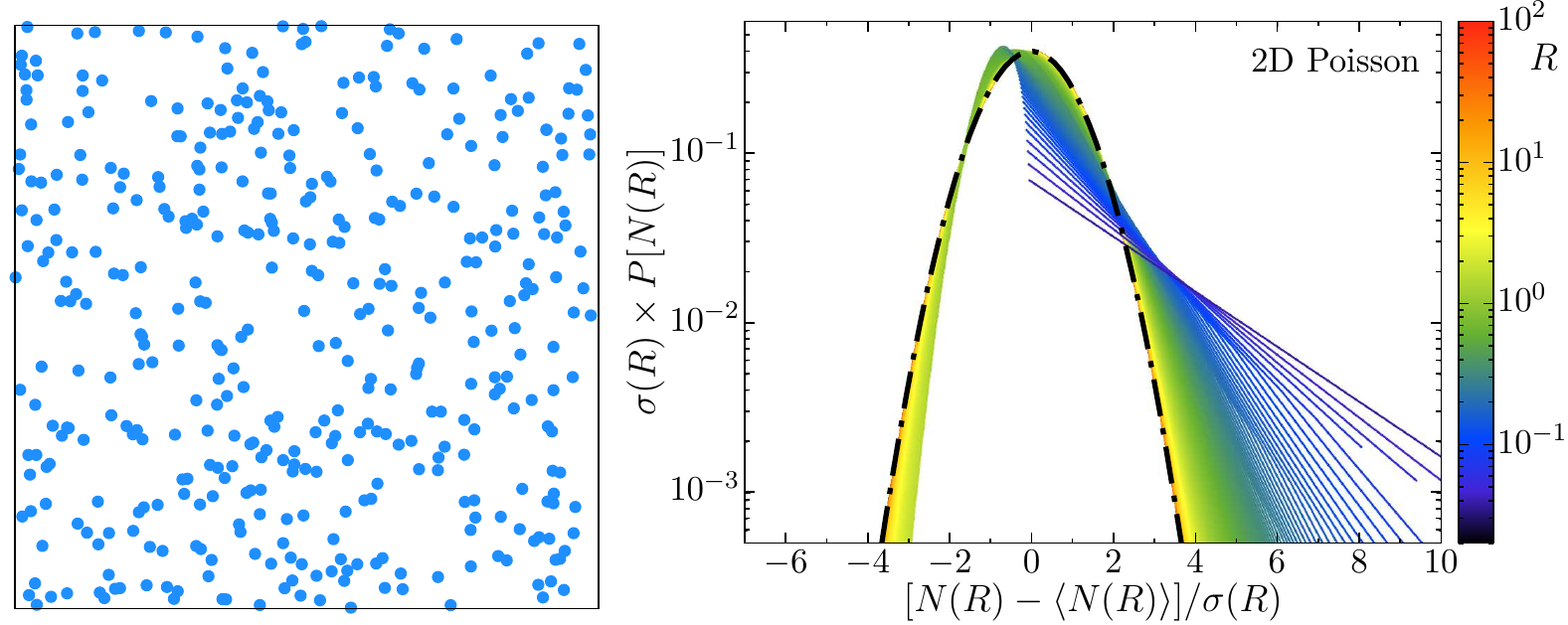}}\hfill%
	\centering{\includegraphics[width = 0.48\textwidth]{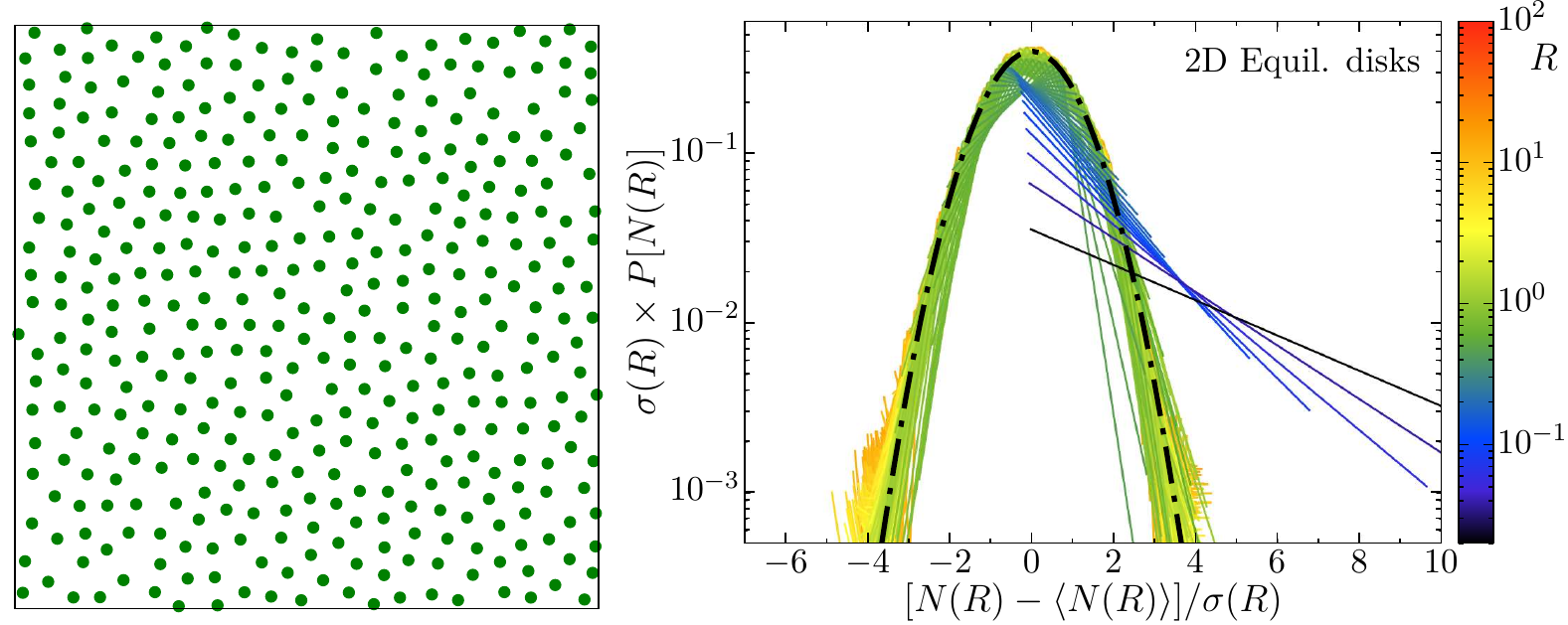}}\\
	\centering{\includegraphics[width = 0.48\textwidth]{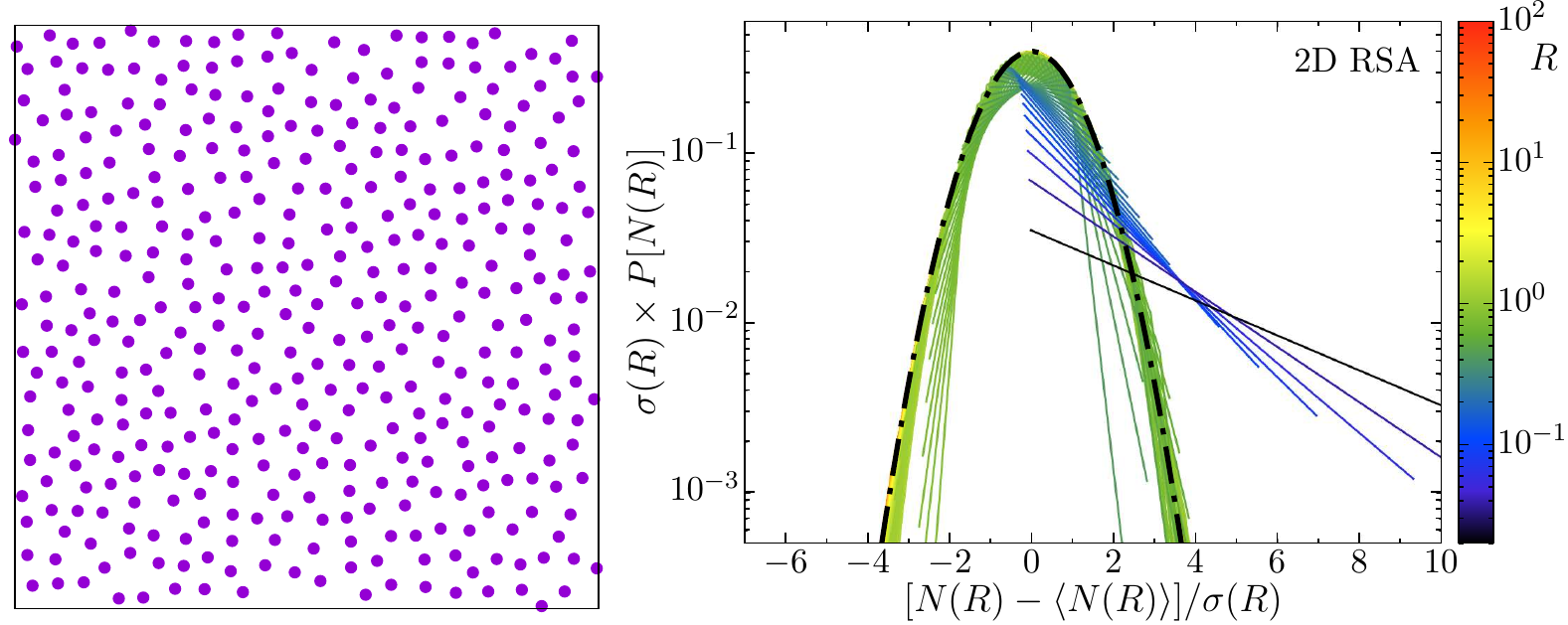}}\hfill \hbox{}
	\caption{Representative images of configurations of each of the five 2D nonhyperuniform models, beginning with the most nonhyperuniform one (HIP) down to 
	one that has the smallest structure factor at the origin (saturated RSA). The corresponding standardized probability distributions for different window
	radii are also included; deepest blue color (darkest shade) corresponds to zero radius and the deep orange color corresponds to the largest considered radius. }
	\label{models-nonhyper}
	\end{figure*}	

For the Poisson and super-Poissonian models (cluster and HIP) across dimensions, both the 
skewness and excess kurtosis decay monotonically; see Fig.~\ref{cumulants}.
By contrast, the skewness and excess kurtosis oscillate about zero for the lattice, 
stealthy, equilibrium and RSA systems across all dimensions because all 
of them exhibit at least short-range order. It should not go unnoticed how the oscillations 
in $\gamma_1 (R)$ and $\gamma_2(R)$ are related to short- or long-range 
order at the level of three- and four-body correlation
functions, $g_3$ and $g_4$, as can be seen from the  
explicit formulas (\ref{C3}) and (\ref{C4}) for the skewness and excess kurtosis.  In the instances of RSA and equilibrium packings, 
oscillations in $\gamma_1 (R)$ and $\gamma_2(R)$ arise from strong 
short-range order exhibited by  $g_3$ and $g_4$.

The disordered hyperuniform systems that we consider, stealthy and URL point processes,
have extraordinary number fluctuation behaviors. For 2D and 3D stealthy systems, both 
$\gamma_1 (R)$ and $\gamma_2(R)$ strongly oscillate about zero; see Fig.~\ref{cumulants}. This suggests
that at the level of $g_3$, stealthy systems, counterintuitively, exhibit significant 
ordering on much larger length scales than the short-range order
seen in the pair correlation function \cite{To15}. For 1D stealthy systems, $\gamma_1 (R)$ and $\gamma_2(R)$
show even stronger oscillations than their higher-dimensional counterparts, indeed
reflecting possible long-range order present in $g_3$ and $g_4$. It is remarkable that the
skewness and excess kurtosis can detect such anomalous long-range order that would
not be expected based solely on the behavior of the pair correlation function. Another reason that supports
the capacity of $\gamma_1 (R)$ and $\gamma_2(R)$ to detect unusual
long-range order is the cloaked URL model, which at the level
of the pair correlation function would be considered to be
highly disordered. Whereas $\gamma_1(R)$ for  this model is a monotonically decreasing 
function of $R$, $\gamma_2(R)$ exhibits oscillations.
This is entirely consistent with the fact that the 
periodicity of the underlying lattice is completely hidden at the level of the
three-point correlation function but manifests itself for the first
time in $g_4$ \cite{Kl20a}.

\subsection{Number Distributions for the Models Across Dimensions}
\label{dist}

\begin{figure*}[t]
	\centering{\includegraphics[width = 0.48\textwidth]{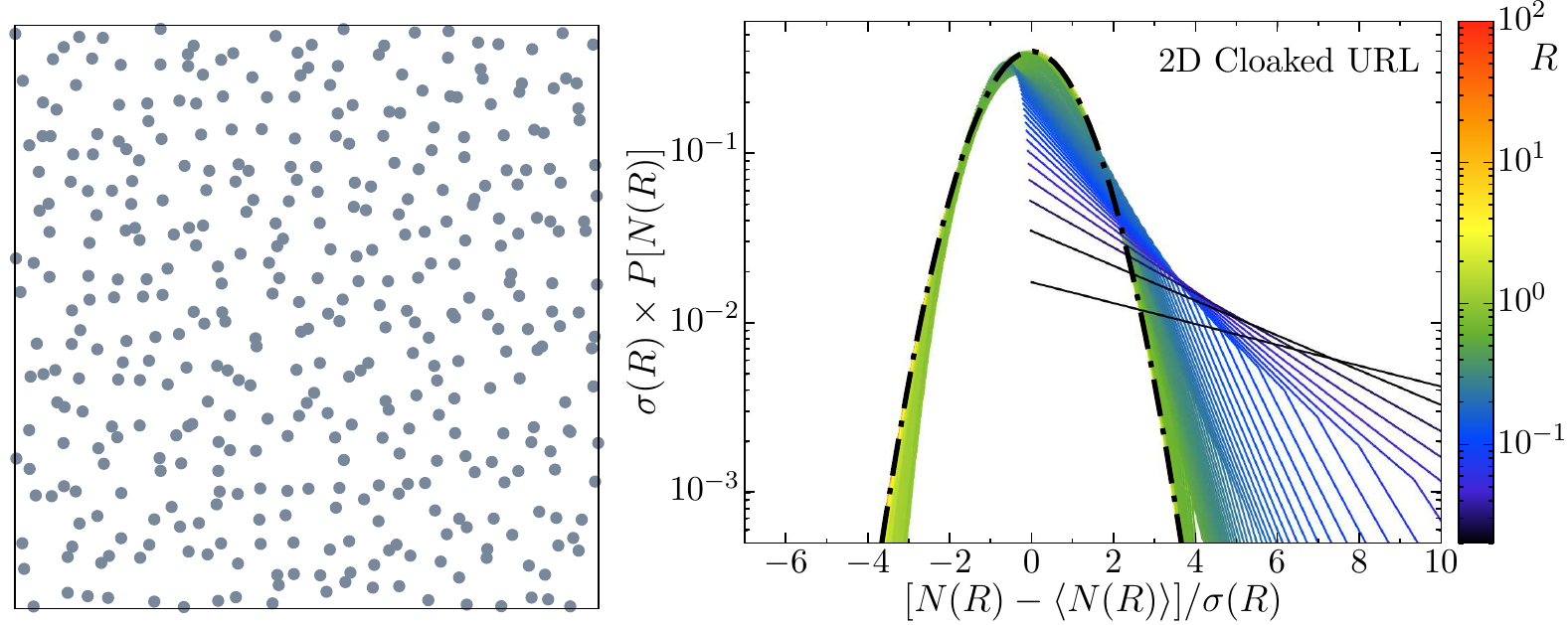}}\hfill%
	\centering{\includegraphics[width = 0.48\textwidth]{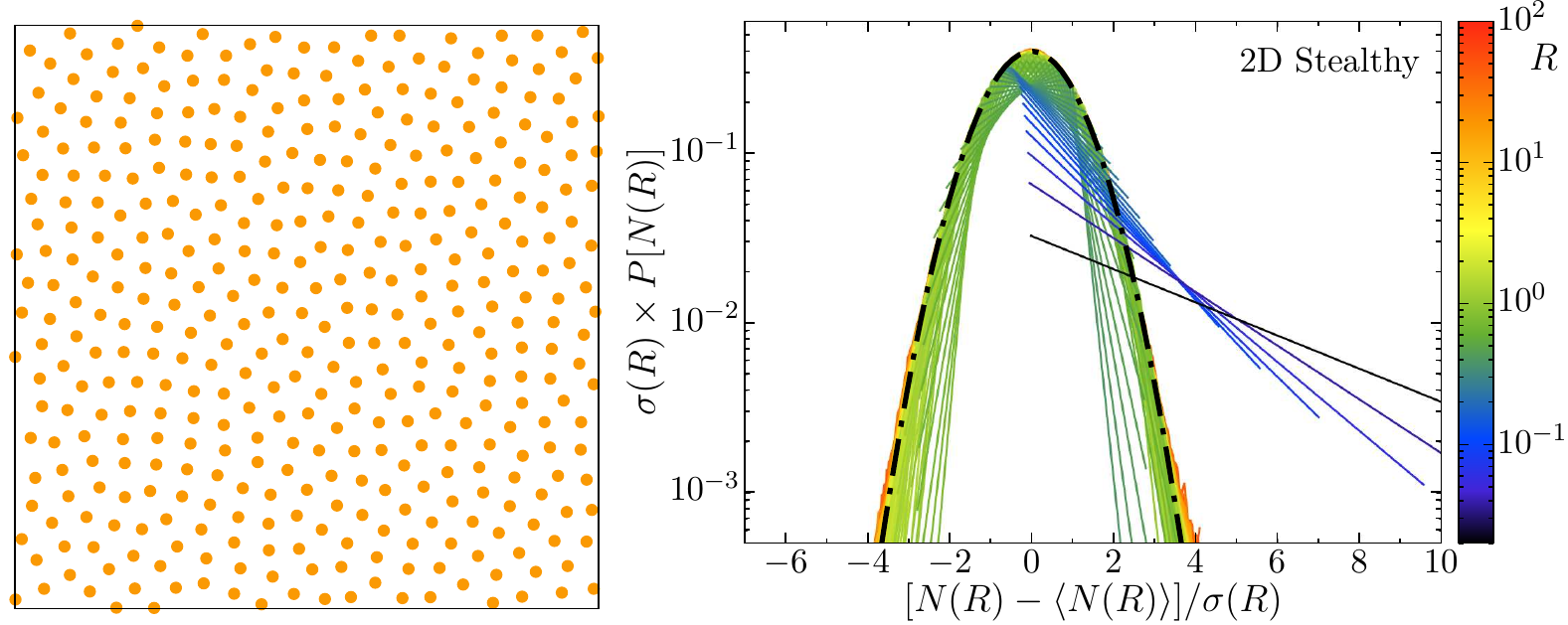}}\\
	\centering{\includegraphics[width = 0.48\textwidth]{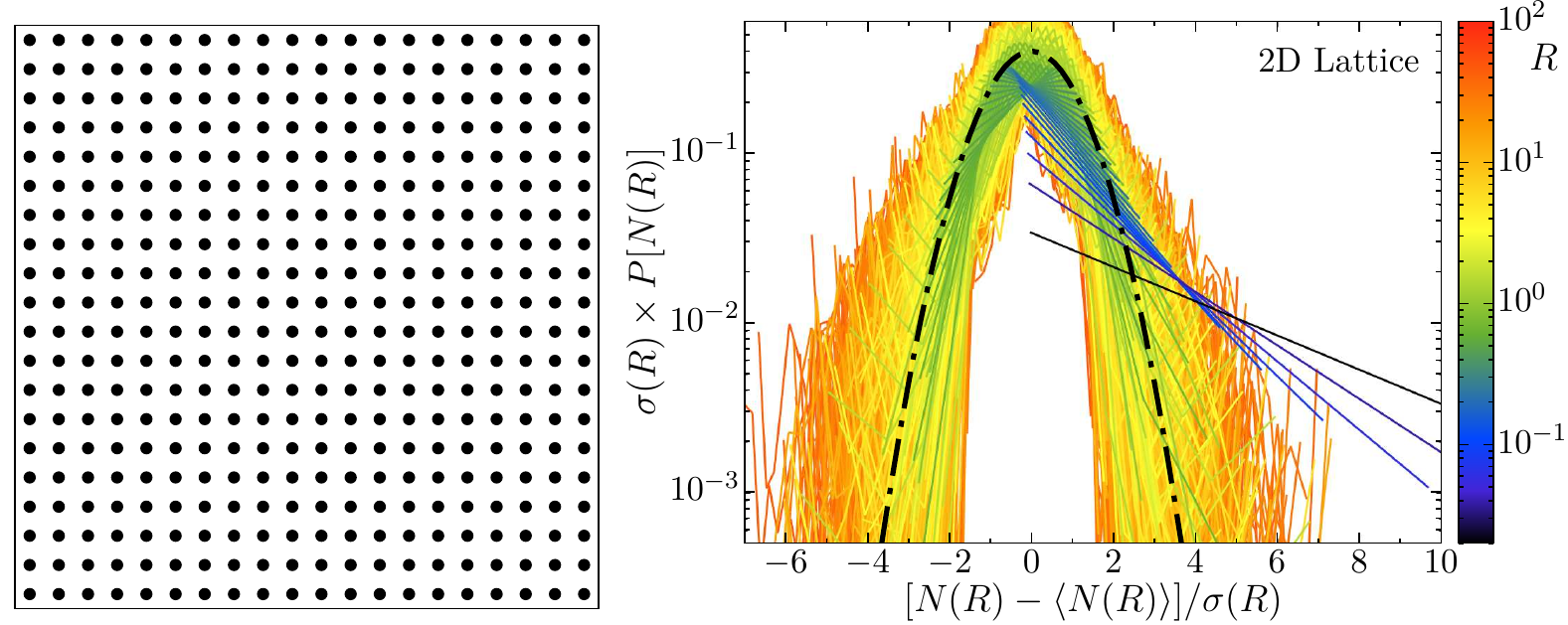}}\hfill \hbox{}
	\caption{Representative images of configurations of each of the three 2D hyperuniform models and their corresponding standardized probability distributions, where 
	the deepest blue color (darkest shade) corresponds to zero radius and the deep orange color corresponds to the largest considered radius.}
	\label{models-hyper}
	\end{figure*}

\begin{figure*}[t]
\centering{\includegraphics[width = 1.0\textwidth]{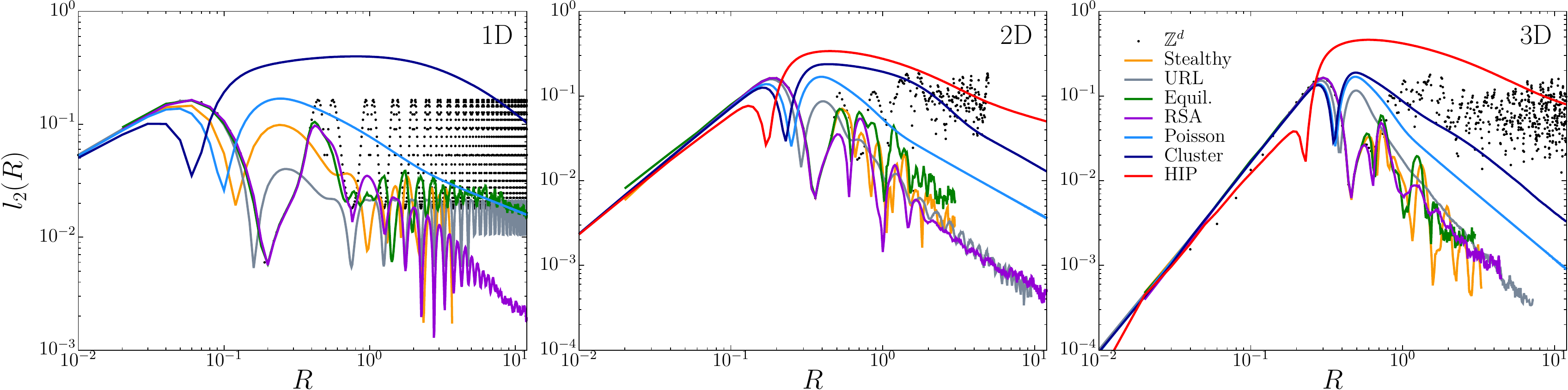}}
\caption{The  Gaussian distance metric $l_2(R)$ versus the window radius $R$ for all considered models across
the first three space dimensions.}
\label{metrics}
\end{figure*}

Figures~\ref{models-nonhyper} and \ref{models-hyper} show representative configurations of all of the 2D models and their corresponding
standardized number distributions. (We provide corresponding figures for all 1D and 3D models in the SM). 
Except for hypercubic lattices for any $d$ and 1D hyperuniform systems of class I,  all of the considered models across dimensions obey a CLT.
It is noteworthy that for all models, except the hypercubic lattices, the number distribution functions $P[N(R)]$ are {\it unimodal}, i.e.,
one with a single peak. (For small radii $R \lessapprox 0.5$, the number distributions are monotonically decreasing, which are still considered to be unimodal.)
Recall that for all models, except the hypercubic lattice and the class I hyperuniform models in one dimension, the skewness and excess
kurtosis tend to zero for large $R$ for all dimensions.
For well-behaved unimodal distributions, such a vanishing of both $\gamma_1(R)$
and $\gamma_2(R)$ indicates a tendency to a CLT.
Indeed, this is confirmed by visual inspection of our corresponding evaluations of the number distributions for all dimensions;
see Figs.~\ref{models-nonhyper} and \ref{models-hyper}.
Our conclusions about the tendencies to CLTs for sufficiently large $R$ are further confirmed by our evaluations
of the Gaussian distance metric $l_2(R)$ for all models across dimensions, which are plotted
in Fig. \ref{metrics}. For $d = 2$ and $d = 3$, one clearly sees that disordered hyperuniform point
processes are better approximated by the normal distribution than their 
nonhyperuniform counterparts at a given large value of $R$. This is consistent with the visual
inspections of the plots presented  Figs. ~\ref{models-nonhyper} and \ref{models-hyper}.
From Fig.~\ref{metrics} and Fig. S8 of the SM, one can ascertain, for each model, 
a radius $R_0$ above which the distribution can be deemed to be approximately Gaussian,
i.e.,  $l_2(R)$ is below some threshold for $R >R_0$, implying that it is
approximately determined by only the mean and the variance. Typically, we find that $R_0$ is orders of magnitude smaller for our hyperuniform and standard nonhyperuniform models than for our super-Poissonian models.

Our results clearly show that the hypercubic lattices across dimensions 
do not obey a CLT; see Figs.~\ref{cumulants},
\ref{models-hyper} and \ref{metrics}.
This is due to the fact that lattice points in general are ``rigid''
in the sense that fluctuations in $N(R)$ are always stringently bounded 
from below and above for any value of $R$ (as discussed in Sec.
\ref{elem}) due to their inherent {\it long-range} order. 
Hence, $P[N(R)]$ is highly sensitive to the value of $R$, so that both 
$\gamma_1(R)$ and $\gamma_2(R)$ rapidly oscillate around zero, but the 
amplitudes of the oscillations do not vanish as $R$ increases. It is already
well-established that the local variance $\sigma^2(R)$ of lattices exhibits such
rapid oscillations; see Ref. \cite{To03a} and references therein.
Because of the rapid  oscillations in
$\gamma_1(R)$, $\gamma_2(R)$, and
$l_2(R)$ for hypercubic lattices, we  represent the 
data by points instead of curves, which would require interpolation 
between the data points.
Note that the corresponding distance metrics $l_2(R)$ are bounded from 
below for any value of $R$.
These observations are consistent with the visual inspections of the 
number distribution shown in Fig.
\ref{models-hyper} for $d=2$ and those for $d=1$ and $d=3$ given in the 
SM.

\subsection{Gamma-Distribution Approximation for All Models Obeying a CLT}
\label{gamma}

\begin{table*}[t]
    \caption{Large-$R$ asymptotic scalings of
    $\sigma^2(R)$, $\gamma_1(R)$, $\gamma_2(R)$ and $l_2(R)$ for all of our  models across dimensions that obey a CLT,
as obtained by the approximation of  $P[N(R)]$ by the gamma distribution function for each model for any $d$.}
    \label{tab:scalings}
    \centering
    \setlength{\tabcolsep}{12pt}
    \begin{tabular}{c |c |c |c}
    \toprule
     Descriptor    & Stealthy       \& URL            & Cluster, Poisson, RSA, \& Equilibrium         & HIP\\
     \colrule
     $\sigma^2(R)$ & $R^{(d-1)}$                     & $R^{d}$                   & $R^{2d-1}$ \\
     $\gamma_1(R)$ & $R^{-(d+1)/2}$                  & $R^{-d/2}$                & $R^{-1/2}$  \\
     $\gamma_2(R)$ & $R^{-(d+1)}$                    & $R^{-d}$                  & $R^{-1}$    \\
     $l_2(R)$      & $R^{-(d+1)/2}$                  & $R^{-d/2}$                & $R^{-1/2}$  \\
     \botrule
    \end{tabular}
\end{table*}

\begin{figure*}[t]
	\includegraphics[width=\textwidth]{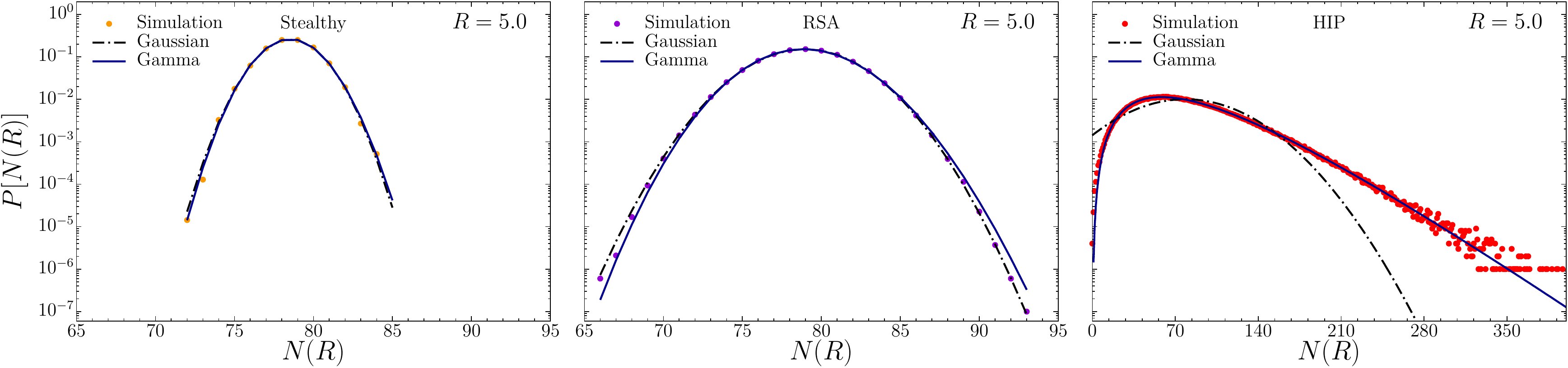}
	\caption{{Comparison of simulation data for the probability distributions $P[N(R)]$  with $R=5$ for 
	three 2D models (stealthy hyperuniform, RSA, and HIP) to corresponding
  gamma-distribution approximations with   the same mean value and variance. 
	The gamma distribution provides good approximations for all models 
	that obey a CLT across the first three dimensions considered, including the ones not shown here.}
	\label{fig:approx}}
  \end{figure*}

We have studied a variety of well-known closed-form probability distributions
to determine those which  best approximate the actual distributions for finite $R$ for all
of our models. Remarkably, we have ascertained that the gamma 
distribution provides a good approximation to the number distribution 
$P[N(R)]$ for all  models that obey a CLT
across all dimensions for intermediate to large values of $R$. (Recall that the hypercubic lattices and 1D hyperuniform systems do not obey
a CLT.)  The gamma distribution is defined by
\begin{equation}
  P[N(R)=m] = \frac{1}{\Gamma(k) \theta^k} m^{k-1}e^{-m/\theta},
  \label{eq:gamma}
\end{equation}
where $k$ and $\theta$ are shape and scale parameters, respectively, 
which are related to the mean and variance of $P[N(R)]$ as 
follows: 
\begin{align}
  \langle N(R) \rangle =& k \theta,\\
  \sigma^2 (R) =& k\theta^2 .
\end{align}
It immediately follows that the associated skewness and excess kurtosis can be expressed
simply in terms of the mean and number variance, yielding
\begin{align}
  \gamma_1(R) =& \frac{2}{\sqrt{k}}= 2 \qty[\frac{\sigma^2(R)}{{\langle N(R) \rangle}^2}]^{1/2}, \\
  \gamma_2(R) =& \frac{6}{k} = 6 \qty[\frac{\sigma^2(R)}{{\langle N(R) \rangle}^2}].
\end{align}
Figure ~\ref{fig:approx} compares the gammma-distribution approximations to simulation
data for  a representative disordered hyperuniform model (stealthy), a standard nonhyperuniform model (RSA)
and an antihyperuniform model (HIP) in two dimensions for $R=5$. Visual inspection reveals the gamma distribution provides a good approximation in each case. Similar good agreement between the gamma-distribution  approximation
and simulation data was found for other models (not shown in Fig. \ref{fig:approx}) obeying a CLT across dimensions, as discussed
in the SM.

\begin{figure*}[t]
	\centering
	  \includegraphics[width=1.0\linewidth]{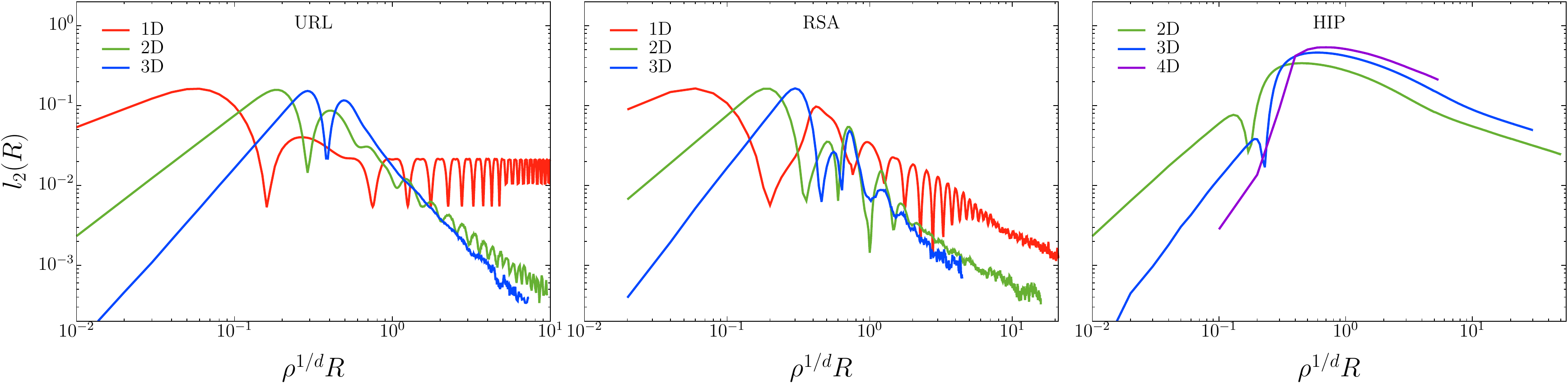}\hfill%
	  \caption{The Gaussian distance metric $l_2(R)$ versus $R$
	for a representative disordered hyperuniform model (URL), sub-Poisson nonhyperuniform model
	(RSA) and anti-hyperuniform model (HIP) across the first three space
	dimensions, respectively.}
	\label{decorrelation}
	\end{figure*}

Importantly,  the approximation of $P[N(R)]$ by a gamma distribution enables us
to estimate  the large-$R$ scalings of $\gamma_1(R)$ and $\gamma_2(R)$ for all models across 
dimensions that obey a CLT. Specifically, we find  $\gamma_1(R)\sim R^{-1/2}\text{~and~}\gamma_2(R)\sim R^{-1}$  for the antihyperuniform HIP,
$  \gamma_1(R)\sim R^{-d/2}\text{~and~}\gamma_2(R)\sim R^{-d}$ for standard nonhyperuniform models, and
$  \gamma_1(R)\sim R^{-(d+1)/2}\text{~and~}\gamma_2(R)\sim R^{-(d+1)}$  for the hyperuniform models. Table \ref{tab:scalings} summarizes these scaling behaviors.
The fact that the excess kurtosis decays to zero faster than the skewness for any model that obeys a CLT, whether hyperuniform or not, implies that the dominant asymptotic correction of the gamma distribution to a CLT for large $R$ is determined by the skewness; see Appendix \ref{asympt} for
a proof.
It is noteworthy that these predictions based on the gamma distribution are 
consistent with numerical findings for all nonhyperuniform systems 
and the antihyperuniform HIP using an independent method that employs 
certain running averages of $\gamma_1(R)$ and  $\gamma_2(R)$; see the SM for 
details. Moreover, for 2D and 3D disordered hyperuniform (stealthy and URL) models, the predictions from the  gamma-distribution approximations
are also consistent with  the observed scalings of the skewness. Due to the strong oscillations
in the excess kurtosis for 2D and 3D stealthy and URL processes described above,
it was numerically difficult to definitively determine their scalings from the running-average
method. It should not go unnoticed that the exact formulas for the skewness and excess kurtosis, 
Eqs. (\ref{gamma1-poisson}) and (\ref{gamma2-poisson}), respectively, of the Poisson distribution
are consistent with the scalings predicted by the gamma-distribution approximation, lending
additional validity to the latter. 
Furthermore, we have found that for all models across dimensions that obey a CLT, the running-average procedure yields scalings for $l_2(R)$ that agree with the corresponding gamma-distribution approximations (see SM), which are identical to the scalings for the skewness $\gamma_1(R)$.
The proof that $l_2(R)$ has exactly the same scaling as $\gamma_1(R)$ for the gamma distribution is given in Appendix \ref{asympt}.
Furthermore, our approximations of $P[N(R)]$ by gamma distributions are also consistent with CLTs, since they converge to the Gaussian distribution in the limit of $k\to\infty$ (i.e, $R \to \infty$),
as described in Appendix \ref{asympt}.

According to Table \ref{tab:scalings},  convergence to a CLT is slowest for the HIP (proportional to $R^{-1/2}$ for $d\ge 2$), 
followed by the Poisson cluster process, and  the Poisson process. RSA and equilibrium packings have the same scaling
behaviors as the Poisson cluster and Poisson processes, but with smaller coefficients of proportionality.
The convergence to a CLT is fastest for the disordered hyperuniform 
processes for $d=2$ and $d=3$~\footnote{
In a less detailed fluctuation study of the ideal gas, equilibrium 
hard-spheres, and MRJ sphere packings, Klatt and Torquato~\cite{Kl16} 
consistently found that the convergence of the number distributions to a 
Gaussian distribution was slowest for the Poisson distribution and 
fastest for hyperuniform MRJ sphere packings.}

\subsection{Effect of Dimensionality on the Approach to a CLT}
\label{decorrelate}

For any particular $d$-dimensional model that eventually obeys a CLT, does the number distribution function $P[N(R)]$
tend to Gaussian-like behavior  faster  as  the space dimension $d$ increases? As noted in the Introduction, this question
can be answered by appealing to the decorrelation principle \cite{To06b}, which
states unconstrained correlations in disordered packings that exist
in low dimensions vanish as $d$ tends to infinity, and all higher-order correlation functions
$g_n$ for $n \ge 3$ may be expressed in terms of the number density $\rho$ and pair correlation function $g_2$.
The decorrelation principle  begins to manifest itself in low dimensions
for disordered packings \cite{Sk06,To06a,To06d} but  other disordered systems with strongly repulsively interacting  particles, including
 fermionic  \cite{To08b} and Gaussian-core point processes \cite{Za08}.
We know that the number distribution function $P[N(R)]$ generally involves certain integrals
over all of the $n$-body correlation functions. Therefore, the decorrelation principle 
implies that for any  $d$-dimensional model that decorrelates with $d$,  $P[N(R)]$ increasingly becomes Gaussian-like as $d$ increases,
since the first and second moments, determined by $\rho$ and $g_2$, dominate the distribution. 
By the same token, for any model that correlates with increasing $d$, $P[N(R)]$ increasingly deviates from the normal distribution
as $d$ increases. We have verified these broad conclusions for the models
studied in this article.  In Fig. \ref{decorrelation}, we plot the Gaussian distance metric $l_2(R)$
for a representative disordered hyperuniform model (URL), a sub-Poisson nonhyperuniform model
(RSA) and an anti-hyperuniform model (HIP) across the first three space
dimensions, respectively.  (Recall that the URL obeys a CLT for $d \ge 2$.) As expected, we see that for both the URL and RSA models for a fixed value
of $R$ (for $R>1$), $l_2(R)$ decreases with increasing $d$ (increasingly tends to Gaussian-like behavior) 
because of decorrelation, while for the HIP, $l_2(R)$ increases with increasing $d$
(moves away from Gaussian-like behavior) because it increasingly correlates with $d$. 
Importantly, when comparing fluctuations across dimensions, one must choose a
meaningful length scale to make the window radius $R$ dimensionless. A simple 
and good choice is $\rho^{1/d}$, which is proportional to the mean
nearest-neighbor distance in space dimension $d$, and explains why the
horizontal axes in each subfigure of Fig. \ref{decorrelation} is $\rho^{1/d} R$.
 
\section{Conclusions and Discussion}
\label{conclusions}

Via theoretical methods and high-precise simulation studies, we  accurately quantified the skewness $\gamma_1(R)$, excess kurtosis $\gamma_2(R)$ 
and the number distribution  $P[N(R)]$ for eight different models of statistically homogeneous point processes
in two and three dimensions:  five nonhyperuniform models, one of which is
anti-hyperuniform (HIP), and three hyperuniform models. Analogous
models were also examined in one dimension, except for HIP, which  is not defined in this dimension.
 We validated our simulation results for  $\gamma_1(R)$, $\gamma_2(R)$ and $P[N(R)]$
for all models by showing that they are in excellent agreement with  rigorous bounds and exact
results that we derived for the applicable ranges of $R$. For all disordered hyperuniform models, our explicit general formulas 
for  $\gamma_1(R)$ and $\gamma_2(R)$  in terms of $n$-body information [Eqs. (\ref{C3}) and (\ref{C4})] enable us to infer
the existence of ``hidden" type of long-range  order that manifests itself for the first time 
at the three-body level or higher. Thus,  the
skewness and excess kurtosis have the capacity to detect anomalous long-range order that would
not be expected based on the behavior of the pair correlation function alone.

We have introduced a novel Gaussian ``distance" metric $l_2(R)$
to ascertain the proximity of  $P[N(R)]$ to the normal distribution for each model as a function of $R$.
We have verified that $l_2(R)$ is a sensitive metric via numerical and theoretical methods. 
Since the distributions for all
models  (except the lattices) across dimensions are unimodal, the tendency
to a CLT corresponds to the skewness and excess kurtosis simultaneously tending to zero.
Almost all of the considered models across dimensions obey a CLT. We found that disordered hyperuniform point
processes are better approximated by the normal distribution than their 
nonhyperuniform counterparts at a given large value of $R$.
We proved that any 1D hyperuniform system of class I as well as the hypercubic lattice for any $d$ cannot obey a CLT.
Similarly, a general lattice in any dimension cannot obey a CLT. 

It is noteworthy that we discovered that the gamma distribution provides a good approximation to the number distribution  $P[N(R)]$ for all  models that obey a CLT
across all dimensions for intermediate to large values of $R$, enabling us  to estimate  
the large-$R$ scalings of $\gamma_1(R)$,  $\gamma_2(R)$ and $l_2(R)$. These predictions were corroborated by corresponding simulation results in almost all instances,
as detailed in the SM. It is only in the special cases of the excess kurtosis for 2D and 3D stealthy and URL processes
where the running-average method was not reliable enough  to definitively determine their scalings due to strong oscillations, and
thus this represents a simulation challenge for future research. Among all models, the convergence to a CLT is generally fastest for the disordered hyperuniform 
processes in two and three dimensions such that $\gamma_1(R)\sim l_2(R) \sim R^{-(d+1)/2}$ and $\gamma_2(R)\sim R^{-(d+1)}$ for large $R$. 
The convergence to a CLT is slower for standard nonhyperuniform models such that 
$\gamma_1(R)\sim  l_2(R) \sim R^{-d/2}\text{~and~}\gamma_2(R)\sim R^{-d}$.
Not surprisingly, the convergence to a CLT is slowest for the antihyperuniform HIP model such that $\gamma_1(R)\sim l_2(R) \sim R^{-1/2}\text{~and~}\gamma_2(R)\sim R^{-1}$.
Using the decorrelation principle,  we elucidated why any $d$-dimensional
model  that ``decorrelates" or ``correlates" with $d$ corresponds to a  $P[N(R)]$ that
increasingly moves toward or away from a CLT, respectively.

It has recently been reported that in a 2D absorbing phase of a lattice 
gas model, the density distribution at a hyperuniform length scale 
increasingly deviates from a Gaussian distribution on approach to the 
critical point~\cite{Zh20c}, corresponding to a class III hyperuniform 
system~\cite{To18a}.
These distributions are distinguished from those of our models in that
they have heavy left tails ($\gamma_1(R) < 0$).
This implies that they cannot be approximated by gamma distributions, as
in the majority of the models studied in the present paper across
dimensions.

A related but distinctly different study from the one considered in the present work concerns 
fluctuations when the window is centered on a point of the point process \cite{To90c,Tr98b}.
It will be interesting in future work to carry out the analogous investigation
of the corresponding skewness, excess kurtosis and number distribution for such particle-based
quantities for hyperuniform and nonhyperuniform models.

\appendix

\section{Asymptotic Behavior of $l_2(R)$ for the Gamma Distribution} \label{asympt}

We prove that the large-$R$ scalings of the distance metric $l_2(R)$ and the skewness $\gamma_1(R)$ are the same for the gamma distribution.
It is important to note that the asymptotic analysis is facilitated by transforming the random variable $N(R)$ to the standardized random variable
$X \equiv (N(R)- \langle N(R)\rangle)/\sigma(R)$. In the large-$R$ limit, the discrete
random variable tends to a continuous one and hence the
continuous-variable analog of the  distance metric $l_2(R)$, defined  Eq. (\ref{l2}),
is given by 
\begin{equation}
  l_2(R) \equiv \left[ \int_{-\infty}^\infty  |F_G(X) - F_\Gamma(X)|^2  dX\right]^{1/2}
\label{L2}
\end{equation}
where $F_G(X)$ and $F_\Gamma(X)$ are the cumulative distribution functions for
the Gaussian and gamma distribution functions, respectively. Specifically,
\begin{equation}
F_G(X)= \frac{1}{2}\left[ 1+ \mbox{erf}(X/\sqrt{2}) \right]
\end{equation}
and 
\begin{equation}
F_\Gamma(X)= 1 - \frac{\Gamma(k,X\sqrt{k}+k)}{\Gamma(k)}
\end{equation}
where $\Gamma(s,y)=\int_{y}^\infty t^{s-1} \exp(-t) dt$ is the upper incomplete gamma function.

\begin{figure}[t]
  \centering
  \includegraphics[width=0.4\textwidth,clip=]{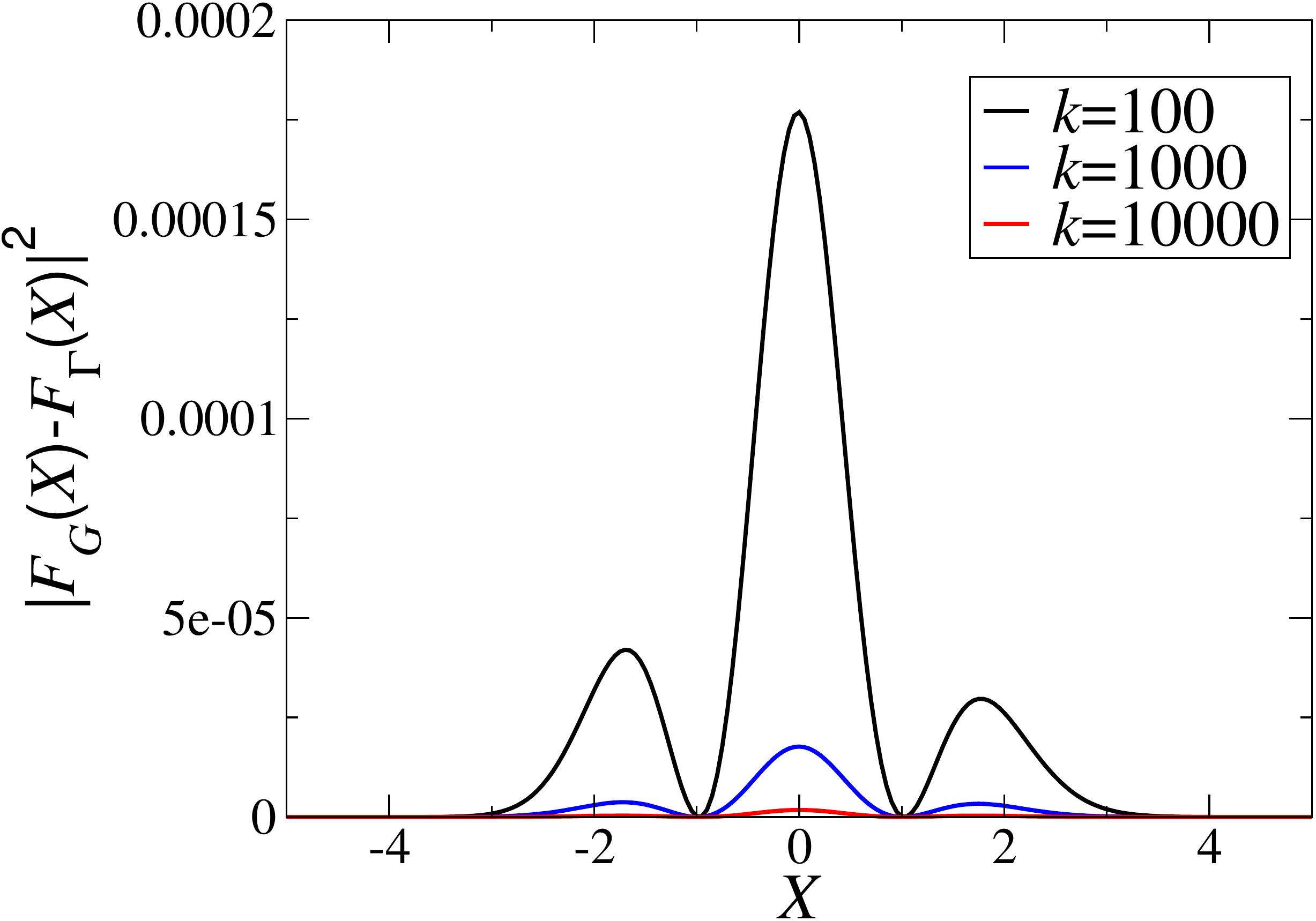}
  \caption{Integrand of (\ref{L2}) versus $X$ for select values of the shape parameter $k$.}
\label{diff}
\end{figure}

Since the $(n+1)$th cumulant of the gamma distribution tends to zero faster than its $n$th cumulant for $n\ge 3$, the greatest deviation of the gamma distribution from the normal one occurs in the vicinity of the origin, i.e., $X=0$, in the large $k$-limit (see Fig. \ref{diff}) and is thus dominated by the asymptotic behavior of the skewness.
Thus, we require the Taylor series expansions of the distributions about $X=0$:
\begin{align} 
F_G(X)=& \frac{1}{2} + \frac{1}{2} \sqrt{\frac{2}{\pi}} X - \frac{1}{12} \sqrt{\frac{2}{\pi}} X^3
+ \frac{1}{80} \sqrt{\frac{2}{\pi}} X^5 \nonumber \\
&- \frac{1}{672} \sqrt{\frac{2}{\pi}} X^7 +{\cal O}(X^9)
\label{gauss}
\end{align}
\begin{align} 
F_\Gamma(X) =& a_0(k) + a_1(k) X +a_2(k) X^2 +a_3(k) X^3 \nonumber \\
& +a_4(k) X^4 + a_5(k) X^5  +a_6(k) X^6 +a_7(k) X^7
\nonumber \\
& +a_8(k) X^8  +{\cal O}(X^9),
\label{Gamma}
\end{align}
where the coefficients $a_m(k)$ ($m=1,2,3,\ldots)$ depend on the shape parameter $k$. 
We know these coefficients explicitly but do not indicate them for reasons of brevity.
The corresponding large-$k$ asymptotic expansions of the first nine coefficients
are 
\begin{eqnarray}
a_0(k) &=& \frac{1}{2} + \frac{1}{6} \sqrt{\frac{2}{k\pi}} +{\cal O}(1/k) \\
a_1(k) &=&  \frac{1}{2} \sqrt{\frac{2}{\pi}} +{\cal O}(1/k) \\
a_2(k) &=& - \frac{1}{4} \sqrt{\frac{2}{k\pi}}  +{\cal O}(1/k^{3/2}) \\
a_3(k) &=& - \frac{1}{12} \sqrt{\frac{2}{\pi}} +{\cal O}(1/k) \\
a_4(k) &=&\frac{5}{48} \sqrt{\frac{2}{k\pi}}  +{\cal O}(1/k^{3/2})\\
a_5(k) &=&  \frac{1}{80} \sqrt{\frac{2}{\pi}} +{\cal O}(1/k) \\
a_6(k) &=&  -\frac{7}{288} \sqrt{\frac{2}{k\pi}} +{\cal O}(1/k^{3/2}) \\
a_7(k) &=& - \frac{1}{672} \sqrt{\frac{2}{\pi}} +{\cal O}(1/k) \\
a_8(k) &=&  \frac{1}{256} \sqrt{\frac{2}{k\pi}} +{\cal O}(1/k^{3/2})
\end{eqnarray}

We see that for odd $m$, the leading-order terms of $a_m$ are exactly the same as
the coefficients multiplying $X^m$ in the series expansion (\ref{gauss}) for the
standard normal distribution. Thus, combining (\ref{gauss}), (\ref{Gamma}) and 
the aforementioned asymptotic expansions yield
\begin{equation}
F_\Gamma(X)-F_G(X)= \frac{f(X)}{\sqrt{k}}  + {\cal O}(\frac{1}{k^{3/2}}),
\label{diff-2}
\end{equation}
where $f(X)$ is a function that is {\it localized} about the origin with the following corresponding Taylor series expansion:
\begin{equation}
f(X)=\sqrt{\frac{2}{\pi}} \left[ \frac{1}{6} -\frac{1}{4} X^2  -\frac{5}{48} X^4  -\frac{7}{288} X^6  -\frac{1}{256} X^8 + {\cal O}(X^{10})\right].
\end{equation}
Substitution of (\ref{diff-2}) into (\ref{L2}) gives the large-$k$ asymptotic expansion
of the Gaussian distance metric, i.e.,
\begin{equation}
l_2(R) = \frac{c}{\sqrt{k}} + {\cal O}(\frac{1}{k^{3/2}}),
\end{equation}
where the square of the constant $c$ is  given by
\begin{equation}
c^2 = \int_{-\infty}^\infty f^2(X) dX.
\end{equation}
Finally, since the skewness $\gamma_1(R)$ scales like $1/\sqrt{k}$, we conclude
that $l_2(R) \sim \gamma_1(R)$ for large values of $R$.
Note that the vanishing of $l_2(R)$ for $R\to\infty$ implies a CLT for the gamma distribution as described in the text.

\smallskip

\begin{acknowledgements}
We thank Steven Atkinson for his configurations of 3D equilibrium hard spheres.
This work was supported in part by the National Science Foundation 
under Award No. DGE-2039656 and by the Princeton University
Innovation Fund for New Ideas in the Natural Sciences.
\end{acknowledgements}


\begin{thebibliography}{80}
\bibitem [{S0()}]{S0}%
	  \BibitemOpen
	  \href@noop {} {}\bibinfo {note} {For equilibrium packings in $\mathbb{R}^d$,
	  $S(0)$ is estimated from the fluctuation-compressibility relation $S(0) =
	  \rho k_B T\kappa_T$, where $k_B$ is the Boltzmann constant, $T$ is
	  temperature, $\kappa_T \equiv \rho^{-1} (\frac{\partial \rho}{\partial p})_T$
	  is the isothermal compressibility, and $p/(\rho k_B T)= G_P(\infty)$ is the
	  reduced pressure.}\BibitemShut {Stop}%
\bibitem [{comm-poisson()}]{comm-poisson}%
	  \BibitemOpen
	  \bibinfo {note} {{Analytic formula (6) is employed.}}\BibitemShut {Stop}%
\bibitem [{comm-integer()}]{comm-integer}%
	  \BibitemOpen
	  \bibinfo {note} {{Analytic formulas (70)-(73) are employed.}}\BibitemShut
	  {Stop}%
\end{thebibliography}

%

\end{document}